\documentclass[fleqn,usenatbib]{mnras}

\usepackage{newtxtext,newtxmath}

\usepackage[T1]{fontenc}

\DeclareRobustCommand{\VAN}[3]{#2}
\let\VANthebibliography\thebibliography
\def\thebibliography{\DeclareRobustCommand{\VAN}[3]{##3}\VANthebibliography}

\usepackage{graphicx}	
\usepackage{amsmath}	
\usepackage{multicol}
\usepackage{bm}
\usepackage{float}
\usepackage{booktabs}
\usepackage{threeparttable}
\usepackage{xcolor}
\usepackage[normalem]{ulem}
\usepackage[export]{adjustbox}

\newcommand{\email}[1]{\mbox{\href{mailto:#1}{#1}}}
\newcommand{\bayesn}{\textsc{BayeSN}}
\newcommand{\numpyro}{\textsc{NumPyro}}

\newcommand{\av}{$A_V$}
\newcommand{\mutext}{$\mu$ }

\DeclareMathOperator{\zltn}{ZLTN}
\DeclareMathOperator{\expon}{Exponential}

\DeclareMathOperator*{\argmin}{arg\,min}

\makeatletter
\newlength{\abovecaptionskip}%
\makeatother

\defcitealias{thorp21}{T21}
\defcitealias{yesbutdiditwork}{Y18}

\title[Variational-BayeSN]{Variational Inference for Acceleration of SN Ia Photometric Distance Estimation with BayeSN}

\author[A.~S.~M.\ Uzsoy et al.]{
Ana Sof\'ia M.\ Uzsoy,$^{1,2,3}$\thanks{E-mail: \email{ana\_sofia.uzsoy@cfa.harvard.edu}}
Stephen Thorp,$^{4,1}$
Matthew Grayling,$^{1}$
and Kaisey S.\ Mandel$^{1,5}$
\\
$^{1}$Institute of Astronomy and Kavli Institute for Cosmology, Madingley Road, Cambridge, CB3 0HA, UK\\
$^{2}$Department of Engineering, University of Cambridge, Trumpington Street, Cambridge, CB2 1PZ, UK\\
$^{3}$Center for Astrophysics | Harvard \& Smithsonian, 60 Garden Street, Cambridge, MA 02138, USA\\
$^{4}$The Oskar Klein Centre, Department of Physics, Stockholm University, AlbaNova University Centre, SE 106 91 Stockholm, Sweden\\
$^{5}$Statistical Laboratory, DPMMS, University of Cambridge, Wilberforce Road, Cambridge, CB3 0WB, UK
}

\date{Accepted XXX. Received YYY; in original form ZZZ}

\pubyear{2023}

\begin{document}
\label{firstpage}
\pagerange{\pageref{firstpage}--\pageref{lastpage}}
\maketitle

\begin{abstract}
Type Ia supernovae (SNe Ia) are standarizable candles whose observed light curves can be used to infer their distances, which can in turn be used in cosmological analyses. As the quantity of observed SNe Ia grows with current and upcoming surveys, increasingly scalable analyses are necessary to take full advantage of these new datasets for precise estimation of cosmological parameters. Bayesian inference methods enable fitting SN Ia light curves with robust uncertainty quantification, but traditional posterior sampling using Markov Chain Monte Carlo (MCMC) is computationally expensive. We present an implementation of variational inference (VI) to accelerate the fitting of SN Ia light curves using the BayeSN hierarchical Bayesian model for time-varying SN Ia spectral energy distributions (SEDs). We demonstrate and evaluate its performance on both simulated light curves and data from the Foundation Supernova Survey with two different forms of surrogate posterior -- a multivariate normal and a custom multivariate zero-lower-truncated normal distribution -- and compare them with the Laplace Approximation and full MCMC analysis. To validate of our variational approximation, we calculate the pareto-smoothed importance sampling (PSIS) diagnostic, and perform variational simulation-based calibration (VSBC). The VI approximation achieves similar results to MCMC but with an order-of-magnitude speedup for the inference of the photometric distance moduli. Overall, we show that VI is a promising method for scalable parameter inference that enables analysis of larger datasets for precision cosmology.
\end{abstract}




\section{Introduction}

\subsection{Scientific Context}
Type Ia supernovae (SNe Ia) are well known as ``standardizable candles'' -- astronomical objects for which there are strong correlations between their absolute brightness and apparent features of their light curves \citep[e.g.][]{rust75, pskovskii77, phillips, tripp98}. They are widely used in precision cosmology, and played a key role in the discovery of dark energy \citep{Riess_1998, Perlmutter_1999}. In the current debate \citep[for a review see e.g.][]{hubble_tension_1, freedman2021} over the value of the Hubble--Lema\^itre constant $H_0$, the SN Ia distance ladder anchored by Cepheid variables \citep{riess22} is in tension with the cosmic microwave background \citep{planck}.

Over the past few decades, higher fidelity and greater volumes of data have been obtained for SNe Ia \citep[e.g.][]{calatololo, riess99, astier06, frieman08, kessler09, hicken12, friedman15, scolnic18, foley18, brout19, des24}. This has allowed the sophistication of SN Ia models to grow from simple scaling relations and templates \citep[e.g.][]{rust75, pskovskii77, phillips, hamuy_templates}, to learned representations of light curves \citep[e.g.][]{LCS, MLCS, MLCS2k2, mandel09, mandel11, burns11} and spectral energy distributions \citep[SEDs; e.g.][]{salt, salt2, SiFTO, SNEMO, salt3, parsnip, thorp21, mandel22, Grayling2024}. 

For as long as SNe Ia have been used as standardizable candles for cosmology, the impact of dust in their host galaxies has been discussed extensively by astronomers \citep[e.g.][]{branch92, sandage93, MLCS, riess96_dust, phillips99, krisciunas00, krisciunas07, mandel11, burns14, mandel17, brout21, popovic21, thorp22, karchev23, karchev24, Grayling2024, thorp2024}. Dust along the line of sight to a supernova or other astrophysical source acts to dim the light from the source. Preferential dimming of bluer wavelengths also causes reddening of the light \citep[see e.g.][for general reviews on astrophysical dust]{dust_review, salim20}. Dust in the Milky-Way galaxy is extensively mapped \citep[e.g.][]{sfd98, galaxy_dust, schlafly2}, so it is possible to correct for its effects. However, the host galaxies of individual SNe Ia are not mapped in this way,  so extinction due to host galaxy dust must be modelled probabilistically. This is one of the primary motivations for the \textsc{BayeSN} hierarchical SED model for SNe Ia \citep{thorp21, mandel22, Grayling2024}. Current implementations of the \textsc{BayeSN} model use the No U-Turn Sampler \citep[NUTS;][]{nuts} to perform full Bayesian inference given a set of SN Ia data. This has been applied to studies of host galaxy dust \citep{thorp21, thorp22, Grayling2024}, SN Ia siblings \citep{ward22}, estimation of the Hubble--Lema\^itre constant \citep{dhawan23}, analyses of new near-infrared (NIR) SN Ia datasets \citep{jones22, peterson23, thorp2024}, and lensed supernova time-delay estimation \citep{pierel24}. A simulation-based inference (SBI) implementation of \textsc{BayeSN} has also been developed recently by \citet{karchev23, karchev24}. To ensure \textsc{BayeSN} remains scalable in anticipation of the large datasets expected from future surveys, we present here a variational implementation of the model for accelerated Bayesian inference of distances from SN Ia light curves. This is built on top of the GPU-accelerated \textsc{BayeSN} code developed by \citet{Grayling2024}.

\subsection{The Need for Speed}
The datasets used in SN Ia cosmology have already been growing substantially over recent years \citep[see e.g.][]{csp2, ztf1, aleo23, des24}. With new, powerful telescopes and space missions observing more astronomical objects than ever before, the field of astrophysics is generally becoming more data-driven. The Vera C.\ Rubin Legacy Survey of Space and Time (LSST) is expected to make over 32 trillion observations in over 20 billion galaxies and stars \citep{lsst_info, ivezic19}, and will be transformative for SN Ia and transient science. The \textit{Nancy Grace Roman Space Telescope} will be similarly transformative for SN Ia cosmology at NIR wavelengths \citep{hounsell18, rose_roman}. Scalable analysis methods are critical to maximizing the scientific output from these projects. Moreover, the ``industry standard'' pipelines and validation frameworks \citep[e.g.][]{snana, bbc, pippin} that will be used in the core SN Ia cosmology analyses for next generation surveys typically require methods that can be rapidly applied to tens to hundreds of thousands of simulated SNe for bias modeling.

Over the past two decades, parameter inference problems in astrophysics and cosmology have often been framed in a Bayesian way, and solved with Markov Chain Monte Carlo (MCMC) samplers and other related numerical methods. The development and uptake of user-friendly interfaces to general purpose sampling algorithms \citep[e.g.][]{cosmomc, multinest, emcee, polychord, dynesty} has greatly facilitated this. While posterior exploration with MCMC remains the gold-standard for Bayesian inference, it is often limited by computational time and resources. 

Variational inference (VI; see \S\ref{vi_overview_section} for an introductory overview) has been shown to improve computational efficiency by orders of magnitude  compared to MCMC when applied to various astrophysical problems \citep[e.g.][]{prabhat, VI_astro_examples, vi_precision_cosmology, vi_starfields}. \citet{desoto24} use VI for parametric light curve fitting for real-time supernova classification, and \citet{sanchez2021amortized} use VI amortized with a neural network to fit SN Ia light curves. \citet{villar2022} also present an amortized inference approach, using masked auto-regressive flows (MAFs; \citealp{papamakarios17}) to approximate the posterior distribution (see e.g.\ \citealp{rezende15}). Alternative variational methods -- particularly variational auto-encoders (VAEs) -- have also been applied effectively to supernovae and other transients. \citet{parsnip} and \citet{villar_vae_classifier} present VAEs to learn latent representations in the context of supernova SED fitting and light curve classification, respectively.
With a VAE, the neural network weights of the encoder and decoder are optimized to learn a latent representation by minimizing the average reconstruction loss across all objects in the training set. In contrast, here we use a pre-trained physically-motivated model (BayeSN) for SN Ia SEDs and light curve data and use VI to approximate the true posterior for each individual SN Ia separately. 

\subsection{Overview of Variational Inference (VI)}
\label{vi_overview_section}

Variational Inference (VI) is a machine learning approach that approximates an intractable posterior that cannot be directly sampled with a simpler ``surrogate'' distribution, also known as a ``guide'' \citep{jordan1999introduction}. The parameters defining this surrogate distribution are then optimized to minimize the difference between the surrogate and true posteriors \citep{jordan1999introduction}. For data $x$, model parameters $\bm{\phi}$, and variational parameters $\bm{\zeta}$, the surrogate posterior $q^*_{\bm{\zeta}}(\bm{\phi})$ is determined as follows:
\begin{equation}
    q^*_{\bm{\zeta}}(\bm{\phi})  = \argmin_{q_{\bm{\zeta}}(\bm{\phi}) \in \mathcal{Q}} \; D_\text{KL} [q_{\bm{\zeta}}(\bm{\phi})\;||\;p(\bm{\phi}|x)]
\end{equation}
where $p(\bm{\phi}|x)$ is the true posterior, $D_\text{KL}$ denotes the Kullback-Leibler divergence \citep{kl_divergence}, and $\mathcal{Q}$ denotes a chosen family of distributions \citep{VI_review}. As the true posterior $p(\bm{\phi}|x)$ is intractable and thus unable to be directly computed, VI maximizes the Evidence Lower Bound \citep[ELBO;][]{jordan1999introduction, wainwright2008graphical, VI_review} optimize the parameters $\bm{\zeta}$ defining the surrogate posterior\footnote{Note that in this work, $\log$ refers to natural logarithm (base $e$), while $\log_{10}$ specifies base 10.}:
\begin{align}
    \text{ELBO}(q_{\bm{\zeta}}(\bm{\phi})) &= \mathbb{E}_q[\log p(\bm{\phi},x)] -  \mathbb{E}_q[\log q_{\bm{\zeta}}(\bm{\phi})] \\
    &= \mathbb{E}_q[\log p(x|\bm{\phi})] - D_\text{KL} [q_{\bm{\zeta}}(\bm{\phi})\;||\;p(\bm{\phi})].
    \label{ELBO}
\end{align}
The parameters $\bm{\zeta}$ (hereafter referred to as the ``variational parameters'') define the surrogate posterior $q_{\bm{\zeta}}(\bm{\phi})$. In practice, these expectations over $q$ are approximated by generating samples (``particles'') of model parameters from the surrogate posterior for a given set of variational parameters and computing the Monte Carlo average over these samples. The variational parameters $\bm{\zeta}$ are optimized with gradient descent using the stochastic variational inference algorithm \citep{hoffman_svi, svi_ppl, svi_vae, bbvi, ADVI}. Validation of variational inference algorithms \citep[e.g.][]{yesbutdiditwork, huggins20} can be conducted using simulation-based calibration \citep[][]{cook06, talts18}, Pareto-smoothed importance sampling \citep[PSIS;][]{vehtari15}, or the Wasserstein distance \citep[][]{kr58, wasserstein}.

\subsection{This Work}
This paper investigates whether VI can be an appropriate replacement for MCMC when using the \textsc{BayeSN} model \citep{mandel22} to estimate the distances to SNe Ia from their light curves. We use stochastic variational inference (SVI) in \numpyro{} \citep{pyro, numpyro} to fit for two different forms of surrogate posterior: a multivariate normal distribution, and a multivariate zero-lower-truncated normal (ZLTN) distribution. We evaluate the performance of the two VI implementations, compared to MCMC and the Laplace approximation, on simulated and real data \citep[from the Foundation survey;][]{foley18, jones19}. We explore the trade-off between computational efficiency and model accuracy and validate the performance of VI in each case \citep[following][]{yesbutdiditwork}. In terms of parameter estimation, both of the tested VI posterior forms perform comparably to MCMC, but with an order of magnitude speedup. We then discuss the potential implications of this new variational implementation of \bayesn{} and when VI should be used to approximate posterior distributions for inference of astrophysical parameters. 

In Section \ref{bayesn_section}, we review the details of the \textsc{BayeSN} model, and in Section \ref{implementation_section} we describe our new VI implementation. In Section \ref{data_section} we describe the real (\S\ref{foundation_section}) and simulated (\S\ref{simulation_section}) data. Our results are presented in Section \ref{results_section}, with additional discussion in Section \ref{discussion_section}, and concluding remarks in Section \ref{conclusions}.

\section{The \textsc{BayeSN} Model}
\label{bayesn_section}
\subsection{Overview}
The full details of the \textsc{BayeSN} model can be found in \citet{mandel22}, with subsequent model developments discussed in \citet{thorp21}, \citet{ward22} and \citet{Grayling2024}. In summary, \textsc{BayeSN} is a hierarchical Bayesian model for the spectral energy distributions (SEDs) of Type Ia supernovae (SNe Ia). 
\textsc{BayeSN} defines a forward model for SN Ia light curves including a fixed spectral template \citep{hsiao07, hsiao09}, a functional principal component-based representation of intrinsic SED variation, a physically-motivated parametric model for host galaxy dust extinction \citep{fitzpatrick99}, and a model for residual spectro-temporal perturbations. Throughout this work, we use the version of the model trained by \citet{thorp21} -- hereafter the \citetalias{thorp21} model -- using data from the Foundation Supernova Survey \citep{foley18, jones19}. These data are described further in Section \ref{foundation_section}.

In the \textsc{BayeSN} model \citep[following][eq.\ 12]{mandel22}, the host dust-extinguished rest-frame SED $S_s(t,\lambda_r)$, of a supernova $s$, is defined by
\begin{multline}
    -2.5\log_{10}\left(\frac{S_s(t,\lambda_r)}{S_0(t,\lambda_r)}\right) = M_0 + W_0(t,\lambda_r) \\ + \delta M_s +  \theta_1^sW_1(t,\lambda_r) + \epsilon_s(t,\lambda_r) + A_V^s\xi(\lambda_r;R_V^{(s)}),
    \label{eq:bayesnsed}
\end{multline}
with reference to a spectral template $S_0(t,\lambda_r)$ \citep{hsiao07, hsiao09}, as a function of rest-frame phase $t$ (relative to the time of maximum $B$-band brightness), and wavelength $\lambda_r$. Here, the supernova-level latent parameters are: a light curve shape parameter/functional principal component score $\theta_1^s$; the $V$-band dust extinction $A_V^s$; a time- and wavelength-dependent residual perturbation function $\epsilon_s(t,\lambda_r)$; and a ``gray'' ($t$- and $\lambda_r$-independent) magnitude offset $\delta M_s$. \bayesn{}'s $\theta_1^s$ parameter correlates with the slope of the SED surface and the width-luminosity relation in the optical \citep{mandel22} and is closely related to the $\Delta m_{15}$ parameter (i.e.\ the decline in magnitude in the 15 days following peak; see discussion in appendix E of \citealp{Grayling2024}). The population-level hyperparameters are: a warping function applied to the mean SED $W_0(t,\lambda_r)$; the first functional principal component $W_1(t,\lambda_r)$; and the \citet{fitzpatrick99} dust law $\xi(A_V; R_V)$. The overall normalization factor is fixed $M_0\equiv-19.5$. The dust law slope parameter can be assumed to be the same for all SNe \citep[as in][]{mandel22, thorp2024}, or allowed to vary on a supernova-by-supernova basis \citep[as in][]{thorp21, thorp22, thorp2024, Grayling2024}. The \textsc{BayeSN} model flux for a supernova $s$ in a passband $b$ is given by Eq.\ 6 in \citet{mandel22}:
\begin{multline}
    f_s(t; b) = (1+z_s)\times10^{0.4(Z - \mu_s)}\\\times \int_{\lambda_r\in b} S_s(t, \lambda_r)\times 10^{-0.4A_{V,s}^\text{MW}\xi^\text{MW}[(1+z_s)\lambda_r]}\mathbb{B}_b[(1+z_s)\lambda_r]\,\lambda_r\,d\lambda_r.
    \label{eq:bayesnflux}
\end{multline}
Here, $\mu_s$ is the SN distance modulus, $z_s$ is the redshift, $A_{V,s}^\text{MW}$ is the Milky Way dust extinction along the SN's line of sight (based on the \citealp{galaxy_dust} dust maps), $\xi^\text{MW}$ is the Milky Way dust extinction law \citep{fitzpatrick99}, $Z\equiv27.5$ is a fixed zero-point (equal to that of SNANA; \citealp{snana}), and $\mathbb{B}_b$ is the normalized transmission function of band $b$. The integral in Eq.\ \ref{eq:bayesnflux} is taken over all rest frame wavelengths observable by band $b$.

In practice the functional parameters in the model, $W_0(t,\lambda_r)$, $W_1(t, \lambda_r)$, and $\epsilon(t,\lambda_r)$, are represented by natural cubic splines\footnote{The \citet{fitzpatrick99} dust law is also defined as a cubic spline.} defined by vectors/matrices of knots $\bm{W}_0$, $\bm{W}_1$, and $\bm{e}_s$. Additionally, the time of maximum may not be perfectly known a priori, meaning a correction factor $\Delta t_\text{max}^s$ may be needed. This gives a set of supernova-level latent parameters $\bm{\phi}_s = (A_V^s, \mu_s, \delta M_s, \theta_1^s, \Delta t_\text{max}^s, \bm{e}_s)$, and population-level hyperparameters $\bm{H} = (\bm{W}_0, \bm{W}_1, R_V, \tau_A, \bm{\Sigma}_e, \sigma_0)$. The latter three hyperparameters specify respectively the population distributions of $A_V^s$, $\bm{e}_s$, and $\delta M_s$:
\begin{align}
    P(A_V^s|\tau_A) &= \expon(A_V^s|\tau_A),\label{eq:avprior}\\
    P(\bm{e}_s|\bm{\Sigma}_e) &= \mathrm{N}(\bm{e}_s|\bm{0}, \bm{\Sigma}_e),\\
    P(\delta M_s|\sigma_0) &= \mathrm{N}(\delta M_s|0, \sigma_0^2).
\end{align}
The prior on $\theta_1^s$ is a unit Gaussian, $P(\theta_1^s) = \mathrm{N}(\theta_1^s| 0,1)$, and we adopt a broad and uninformative prior on distances $P(\mu_s) = \mathrm{N}(\mu_s|\mu_{\Lambda\text{CDM}}(z_s), 5^2)$, where $\mu_{\Lambda\text{CDM}}(z_s)$ is the estimated distance under a flat $\Lambda$CDM cosmology. In this work, we adopt a Gaussian prior on the time of maximum correction\footnote{This choice implies a 95\% prior probability that the original maximum date was estimated to within $10\times(1+z_s)$ observer-frame days of the truth.}, $P(\Delta t^s_\text{max})=\mathrm{N}(\Delta t^s_\text{max}|0, 5^2)$. In this work, we are interested in \textsc{BayeSN} as a distance estimator (c.f.\ \citealp{thorp21} \S4.6; \citealp{mandel22} \S2.8) -- i.e.\ given a supernova's photometry, what is the posterior on $\mu_s$ (and the other latent parameters in $\bm{\phi}_s$). For this reason, we will fix the values of the hyperparameters to their posterior mean values, $\hat{\bm{H}}$, estimated by \citet{thorp21} from the full Foundation dataset \citep{foley18, jones19}\footnote{These hyperparameter estimates can be found in the \texttt{BAYESN.T21} directory at \url{https://github.com/bayesn/bayesn-model-files}}. For a supernova $s$, we will have $N$ observed fluxes $\hat{\bm{f}}_s = (\hat{f}_{s,1},\dots,\hat{f}_{s,N})^\top$, with corresponding observation dates\footnote{These are Modified Julian Dates (MJD) in observer-frame days.} $\hat{\bm{T}}_s = (\hat{T}_{s,1},\dots,\hat{T}_{s,N})^\top$, measurement errors $\hat{\bm{\sigma}}_s=(\hat{\sigma}_{s,1},\dots,\hat{\sigma}_{s,N})^\top$, and rest-frame phases $\hat{\bm{t}}_s = (\hat{\bm{T}}_s - \hat{T}_\text{ref}^s)/(1+z_s)$. The phases are defined with reference to an estimated date of maximum light $\hat{T}^s_\text{ref}$. The posterior distribution of $\bm{\phi}_s$ will be given by
\begin{multline}
    P(\bm{\phi}_s|\hat{\bm{f}}_s; \hat{\bm{H}}) \propto P(\bm{\phi}_s|\hat{\bm{H}})
    \\ \times \prod_{n=1}^{N} \mathrm{N}\big(\hat{f}_{s,n}| f(\hat{t}_{s,n} + \Delta t_\text{max}^s;b_n,\bm{\phi}_s, \hat{\bm{H}}), \hat{\sigma}_{s,n}^2\big),
\end{multline}
where 
\begin{multline}
    P(\bm{\phi}_s|\hat{\bm{H}}) = P(\mu_s)\times P(\theta_1^s)\times P(\Delta t_\text{max}^s) \\
    \times P(A_V^s|\tau_A)\times P(\bm{e}_s|\bm{\Sigma}_e)\times P(\delta M_s|\sigma_0),
\end{multline}
the passband for observation $n$ is denoted by $b_n$, $\Delta t_\text{max}^s$ is a correction for the time of maximum\footnote{Rather than fitting for the date of maximum $T_\text{max}^s$ directly, we adopt a parameterization that is defined with reference to a prior estimate $\hat{T}^s_\text{ref}$. The correction factor is related to $T_\text{max}^s$ by: $\Delta t_\text{max}^s=(T^s_\text{max} - \hat{T}^s_\text{ref})/(1+z_s)$.},  and the model flux $f(\hat{t}_{s,n};b_n,\bm{\phi}_s, \hat{\bm{H}})$ is evaluated using Eq.\ \ref{eq:bayesnsed} and $\ref{eq:bayesnflux}$. This posterior distribution is the target of the variational approximation we develop in the subsequent sections of this paper.

\subsection{The Challenge of Dust Extinction}
\label{section:dust_challenge}
The dust extinction parameter $A_V$ can take on any value in the range [0,$\infty$), and can be correlated with other parameters in the posterior distribution. The exponential prior in Equation \ref{eq:avprior} is physically motivated, and favors low values\footnote{The popular choice of an exponential prior (Eq.\ \ref{eq:avprior}) on dust extinction was originated by \citet{MLCS2k2}, motivated by the fact that viewing angles that produce high extinction must be rarer. Theoretical studies \citep{hatano98, commins04, riello05} had also favored extinction values following an exponential distribution for late-type SN Ia hosts (see also \citealp{holwerda15}). This assumption has been scrutinized by several recent works \citep{wojtak23, birdsnack}.}. For SNe Ia with little-to-no dust extinction ($A_V$ close to zero), there can be considerable nonzero posterior density at $A_V\approx0$. However, negative values of $A_V$ are unphysical, and should not be permitted. This presents a challenge for variational inference, as we need an approximate posterior with the following properties:
\begin{enumerate}
    \item $A_V$ constrained positive, all other parameters unbounded;
    \item potential correlation between all parameters (i.e.\ ``full rank''); and
    \item non-negligible density allowed for $A_V$ values arbitrarily close to zero.
\end{enumerate}

A simple solution which satisfies points (i) and (ii) is to fit a multivariate normal to the joint posterior distribution of $\log A_V$ and all other parameters. This is one of the approximations we will consider in subsequent sections of this paper. The disadvantage of this approach is that the the approximate posterior marginal on $A_V$ will be effectively log-normal, tending towards zero density as $A_V\to0$. Underestimating the posterior density at $A_V=0$ can potentially lead to overestimation of extinction in the low-dust limit. A multivariate normal across all parameters (including $A_V$) would satisfy points (ii) and (iii), but would allow for potentially problematic negative values of $A_V$. Points (i) and (iii) could be satisfied simultaneously by using a variational surrogate posterior that is a product of an independent truncated normal distribution on $A_V$, and a multivariate normal for all other parameters. To satisfy all three desiderata, we propose using a variational guide where a zero-lower-truncated normal (ZLTN) distribution approximates the marginal posterior on $A_V$, and the \emph{conditional} distribution of the remaining parameters given $A_V$ is multivariate Gaussian (the exact form is defined in \S\ref{zltn_section}). For $A_V$ far away from zero, the ZLTN distribution recovers the same behavior as a non-truncated Gaussian distribution, but in the low-$A_V$ limit the behavior is more desirable. Using this distribution allows for a more flexible posterior approximation, while still providing the computational speedup of VI over MCMC.

\section{Variational Inference Implementation}
\label{implementation_section}
In our VI implementation, we aim to fit 28 \bayesn{} model parameters: $A_V, \mu+\delta M, \theta_1, \Delta t_\text{max}$, and $\bm{e}$, where $\bm{e}$ is 24 entries for the residual realization. For identifiability reasons, we treat the sum of $\mu$ and $\delta M$ as a single parameter, and separate them in post-processing.  For each posterior form, we provide first a theoretical overview followed by the implementation details. In general, the \bayesn{} model was implemented using the \numpyro{} probabilistic programming language. Within \numpyro{} we use the {\tt SVI} class to perform Stochastic Variational Inference and use the {\tt AutoGuide} functionality to define the form of the surrogate posterior, and constrain the variational parameters $\bm{\zeta}$. 

\subsection{The Laplace Approximation}

The Laplace Approximation approximates a posterior probability distribution as Gaussian centered around the maximum a posteriori (MAP) estimate. Here we approximate the posterior distribution over the parameter vector $\bm{\phi} = \{\log A_V, \mu + \delta M, \theta_1, \Delta t_\text{max}, \bm{e}\}$ using the Laplace approximation, where a log transform is used to enforce a positivity constraint on $A_V$. For some multivariate posterior distribution $p(\bm{\phi}| x)$, with parameters $\bm{\phi}$ and data $x$, we can perform a second-order Taylor expansion about the MAP estimate $\hat{\bm{\phi}}$:

\begin{equation}
    \log p(\bm{\phi}| x)  
    \approx \log p(\hat{\bm{\phi}}| x) - \frac{1}{2} (\bm{\phi} - \hat{\bm{\phi}})^\top \bm{\Xi}\; (\bm{\phi} - \hat{\bm{\phi}}).
\end{equation}
Here $\nabla{\log p(\bm{\phi}| x)} \bigr\rvert_{\bm{\phi} = \hat{\bm{\phi}}} = 0$ by the definition of MAP, and $\bm{\Xi} = - \nabla \nabla{\log p(\bm{\phi}| x)} \bigr\rvert_{\bm{\phi} = \hat{\bm{\phi}}}$ is the Hessian of the negative log-posterior at $\hat{\bm{\phi}}$ \citep[see e.g.][]{Rasmussen_Williams, gelman_book}. The posterior distribution can thus be approximated by a Gaussian distribution,
\begin{equation}
    q_\zeta(\bm{\phi}) \approx \mathrm{N}(\bm{\phi} |\, \hat{\bm{\phi}}, \bm{\Xi}^{-1}).
\end{equation}

The Laplace Approximation was implemented using the {\tt AutoLaplaceApproximation} class within \numpyro{} with parameters initialized to the median of their respective priors. The {\tt AutoLaplaceApproximation} interface is the same as all other {\tt AutoGuides} in \numpyro, but the ELBO is maximized between the true posterior and a Delta function guide to find the MAP estimate. 

\subsection{The Multivariate Normal Distribution}
Here we define the parameter vector $\bm{\phi} = \{\log A_V, \mu + \delta M, \theta_1, \Delta t_\text{max}, \bm{e}\}$ (again using a log transform to constrain $A_V > 0$) and approximate the joint posterior with a surrogate of the form:

\begin{equation}
    q_\zeta(\bm{\phi}) = \mathrm{N}(\bm{\phi} | \bm{\mu}, \bm{\Sigma})
\end{equation}
where variational parameters $\bm{\zeta} = (\bm{\mu}, \bm{\Sigma})$ with mean vector $\bm{\mu}$ and covariance matrix $\bm{\Sigma}$. The multivariate normal guide with a full covariance matrix was implemented using the {\tt AutoMultivariateNormal} class within \numpyro.  

\subsection{The Multivariate Zero-Lower-Truncated Normal Distribution}
Here we define the generic multivariate Zero-Lower-Truncated Normal (MVZLTN) distribution as a multivariate distribution with the first variable being characterized by a ZLTN distribution (as defined below) and the remainder following a multivariate normal distribution conditional on the first variable. As described in Section~\ref{section:dust_challenge}, we use this distribution to impose a positivity constraint on $A_V$ while maintaining a full covariance matrix between all model parameters $\bm{\phi} = \{A_V, \mu + \delta M, \theta_1, \Delta t_\text{max}, \bm{e}\}$. To allow for nonzero density at $A_V=0$, we are not imposing a log transform on $A_V$.

\subsubsection{1D ZLTN Distribution}
If a generic one-dimensional truncated normal distribution is truncated on the left at $\phi=a$ and on the right at $\phi=b$, then the Zero-Lower-Truncated Normal (ZLTN) distribution has $a=0$ and $b=\infty$. The probability density function (PDF) has the form
\begin{equation}\label{eq:sample_zltn}
    \zltn(\phi|\,\mu, \sigma^2) = \begin{cases} \dfrac{1}{\sigma}\dfrac{\psi([\phi-\mu]/\sigma)}{\Psi(\mu/\sigma)} \quad &\text{if }\phi\geq0\\
    0 &\text{otherwise},
    \end{cases}
\end{equation}
where $\mu$ and $\sigma$ are the mean and standard deviation of the untruncated distribution, and $\psi(\phi)$ and $\Psi(\phi)$ are respectively the PDF and cumulative density function (CDF) of a standard unit normal distribution.

\subsubsection{Multivariate ZLTN Distribution}
\label{zltn_section}
We define an $N$-dimensional distribution over $\bm{\phi} = (\phi_t,\phi_u^1,\dots,\phi_u^{N-1})^\top$, characterized by a mean vector $\bm{\mu}$ and a covariance matrix $\bm{\Sigma}$. The first variable ($\phi_t$) has positive support and is ZLTN distributed. We call this the truncated variable. The remaining $N-1$ variables, $\bm{\phi}_u=(\phi_u^1,\dots,\phi_u^{N-1})^\top$, are untruncated and follow a multivariate normal distribution conditional on $\phi_t$. The mean and covariance parameters can be partitioned as:
\begin{equation}
    \bm{\mu} =  \begin{bmatrix}
           \mu_{t} \\
           \bm{\mu}_{u} 
         \end{bmatrix} 
         , \bm{\Sigma} = 
         \begin{bmatrix}
           \sigma_{t}^2 & \bm{\Sigma}_{ut}^\top \\
          \bm{\Sigma}_{ut} & \bm{\Sigma}_{uu}
         \end{bmatrix}
         \label{eq:zltn_params}
\end{equation}
where $\mu_t$ is a scalar, $\bm{\mu}_u$ is a vector of length $N-1$, $\sigma_{t}$ is a scalar, $\bm{\Sigma}_{ut}$ is a vector of length $N - 1$, and $\bm{\Sigma}_{uu}$ is an $(N - 1) \times (N - 1)$ matrix. Here $\mu_t$ and $\sigma_{t}^2$ are parameters of a ZLTN distribution, and the rest are parameters of a multivariate normal distribution. 

First, we define the marginal probability density of the truncated parameter $\phi_t$ as $\zltn(\phi_t|\,\mu_t, \sigma_{t}^2)$. The conditional distribution of the remaining parameters is then:
\begin{equation}
\label{eq:new_conditional_dist}
    \bm{\phi}_u |\, \phi_t \sim \mathrm{N}(\bm{\mu}_{u|t}, \bm{\Sigma}_{uu|t})
\end{equation}
where the conditional mean is
\begin{equation}
    \bm{\mu}_{u|t} = \bm{\mu}_u + \bm{\Sigma}_{ut}\frac{\phi_t - \mu_t}{\sigma_{t}^2}
    \label{eq:conditional_mean}
\end{equation}
and the conditional covariance is
\begin{equation}
    \bm{\Sigma}_{uu|t}  = \bm{\Sigma}_{uu} - \frac{1}{\sigma_{t}^2}\bm{\Sigma}_{ut}\bm{\Sigma}_{ut}^\top
    \label{eq:conditional_cov}
\end{equation}
The log probability density of a full $N$-dimensional parameter vector $\bm{\phi}=(\phi_t,\bm{\phi}_u)^\top$ can be calculated as the sum of the log probability density of the truncated parameter, and that of all other variables conditioned on the truncated one:
\begin{equation}\label{eqn:logq_mzltn}
    \begin{split}
    \log&\,\text{MVZLTN}(\bm{\phi} |\, \bm{\mu}, \bm{\Sigma})\\ &\equiv \log \mathrm{N}(\bm{\phi}_u |\,\bm{\mu}_{u|t}, \bm{\Sigma}_{uu|t}) + \log \zltn(\phi_t|\mu_t, \sigma_{t}^2)
    \end{split}
\end{equation}
The first term on the right hand side is calculating the log probability density of the multivariate Gaussian distribution in Equation~\ref{eq:new_conditional_dist} and the second term is the log probability density of a 1D ZLTN distribution (Equation~\ref{eq:sample_zltn}). Useful marginal probability densities over subvectors of $\bm{\phi}$ are derived in Appendix \ref{zltn_marginals}.

\subsubsection{Using Multivariate ZLTN as a Surrogate}

In our application, we use the MVZLTN distribution constructed above, with a joint density computed as in Eq. \ref{eqn:logq_mzltn}, as the surrogate posterior distribution:
\begin{equation}
q_{\bm{\zeta}}(\bm{\phi}) = \text{MVZLTN}(\bm{\phi} |\, \bm{\mu}, \bm{\Sigma})
\end{equation}
where $\phi_t = A_V$ is the parameter whose distribution is truncated to be positive. The variational parameters are $\bm{\zeta} = (\bm{\mu}, \bm{\Sigma})$ as defined in Equation~\ref{eq:zltn_params}. 

\subsubsection{Custom AutoGuide Implementation}

We first extend the {\tt Distribution} class in \numpyro{} to create a custom joint MVZLTN distribution with the first variable following a ZLTN distribution and all other variables a multivariate normal distribution conditional on the first . We override the {\tt sample()} and {\tt log\_prob()} methods and implement new ones using the equations above. We then extend the {\tt AutoContinous} guide class and create a new type of {\tt AutoGuide} class that has a MVZLTN posterior. Utilizing the {\tt AutoGuide} functionality in \numpyro{} enables us to fit a MVZLTN posterior to any model.

\section{Data}
\label{data_section}
To test our variational inference implementation of \textsc{BayeSN}, we fit both real and simulated SN Ia light curves.

\subsection{Foundation Supernova Survey}
\label{foundation_section}
As our test dataset, we use the first data release of the Foundation Supernova Survey \citep{foley18, jones19}. These data include $griz$ ($\sim3500$--9500~\AA) light curves of 180 spectroscopically normal SNe Ia observed on the Pan-STARRS telescope. The light curves have a fairly homogeneous time sampling, with observations following a nominal sequence of $(-5, 0, 5, 10, 18, 26, 35)$~days relative to peak brightness \citep{foley18}. The filter set, cadence, and telescope used make these data a good representation of other current datasets to which a scalable \textsc{Variational-BayeSN} model will be directly applicable (e.g.\ the Young Supernova Experiment; \citealp{aleo23}; and Pan-STARRS Medium Deep Survey; \citealp{rest14, scolnic18, villar20}). They are also a valuable precursor to the data expected from LSST. Additionally, previous \textsc{BayeSN} analyses \citep{thorp21, ward22} have made use of the Foundation data, and they form the training set of the \citetalias{thorp21} model used throughout this work. We limit ourselves here to the 157 Foundation SNe Ia used in \citet{thorp21}.

\subsection{Simulated Data}
\label{simulation_section}
As well as analyzing real data from Foundation, we also perform tests using data simulated from the \citetalias{thorp21} \textsc{BayeSN} forward model. These data are simulated to closely mimic the properties of the real Foundation data, but also give us access to ground truth parameter values. To simulate our data we sampled parameter values from the following priors\footnote{We use the notation $\text{Exponential}(\tau)$ to denote an exponential distribution with a scale parameter $\tau$. Here $\tau_A$ is a population mean $A_V$.}:
\begin{align}
    A_v &\sim \text{Exponential}(\tau_A=0.194)\\
    \theta_1 &\sim \text{N}(0,1)\\
    z &\sim \text{U}(0.015, 0.08)\\
    \Delta t_\text{max} &\sim \text{N}(0, 5^2)
\end{align}
where the distance modulus $\mu$ was calculated from the sampled redshift value $z$ (assuming a flat $\Lambda$CDM cosmology with $H_0 = 73.24~\text{km}\,\text{s}^{-1}\,\text{Mpc}^{-1}$, $\Omega_0=0.28$, $\Omega_\Lambda=0.72$; \citealp{riess2016}). We then used the \bayesn{} forward model to simulate SEDs and ``observed'' light curves using these sampled parameters, with simulated photometric errors of 0.05~mag, and a Gaussian error of 0.001 applied to the redshifts to mimic spectroscopic redshift error. In our fits, we fix redshift to this ``observed'' value, rather than the true $z$ (reflecting the way in which spectroscopic redshifts are treated in real analyses). In practice, this error only affects the redshift and time dilation of the SED model, and the effect is negligible at low-$z$ after integration through the bandpasses and the addition of photometric error.

The simulations followed an idealized version of the Foundation survey strategy, with a cadence of 4 rest-frame days, with the earliest observation at 8 rest-frame days before peak, and the latest at 36 rest-frame days after peak in $griz$ (with observations in the four bands being coincident in time). The nominal observing sequence in the Foundation survey \citep{foley18} was $[-5, 0, 5, 10, 18, 26, 35]$~days relative to peak (in the observer frame), meaning our simulations cover a comparable range of phases and have similar sampling close to the peak of the light curve. Our assumed photometric uncertainties of $0.05$~mag are comparable to the mean photometric uncertainty of $0.045$~mag for the full \citet{foley18} dataset.

\begin{figure*}
    \centering
        \includegraphics[width=\textwidth]{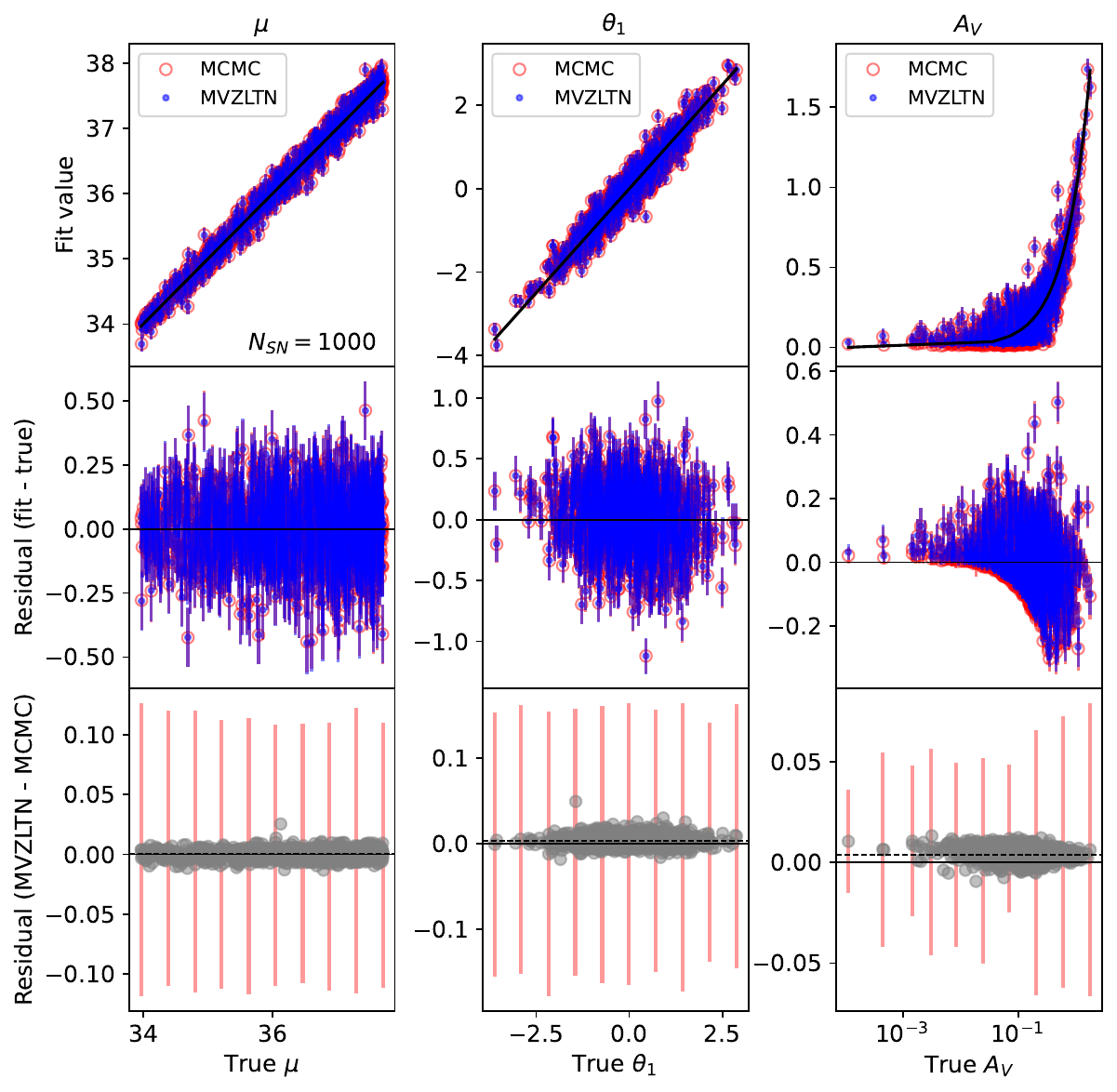} 
        \caption{A comparison of MVZLTN VI posterior fits (blue) and MCMC posterior fits (red) with the true parameter values for 1000 simulate SN Ia for $\mu$, $\theta_1$, and \av. The top row compares the median posterior value vs. the true parameter value, the middle row shows the residuals (fit - true) for MCMC and MVZLTN VI, and the bottom row the residuals (VI - MCMC) as a function of the true parameter value. Black dotted lines in the bottom row denote the median residual (MVZLTN - MCMC) value.
        The red error bars in the bottom row indicate the typical posterior standard deviation derived from MCMC fits to SNe at a set of 10 representative values spanning each parameter range. Note that the true \av{} is plotted on a $\log_{10}$ scale to better show the behavior at low \av.}
        \label{fig:zltn_sims}
\end{figure*}

\begin{figure*}
    \centering
        \includegraphics[width=\textwidth]{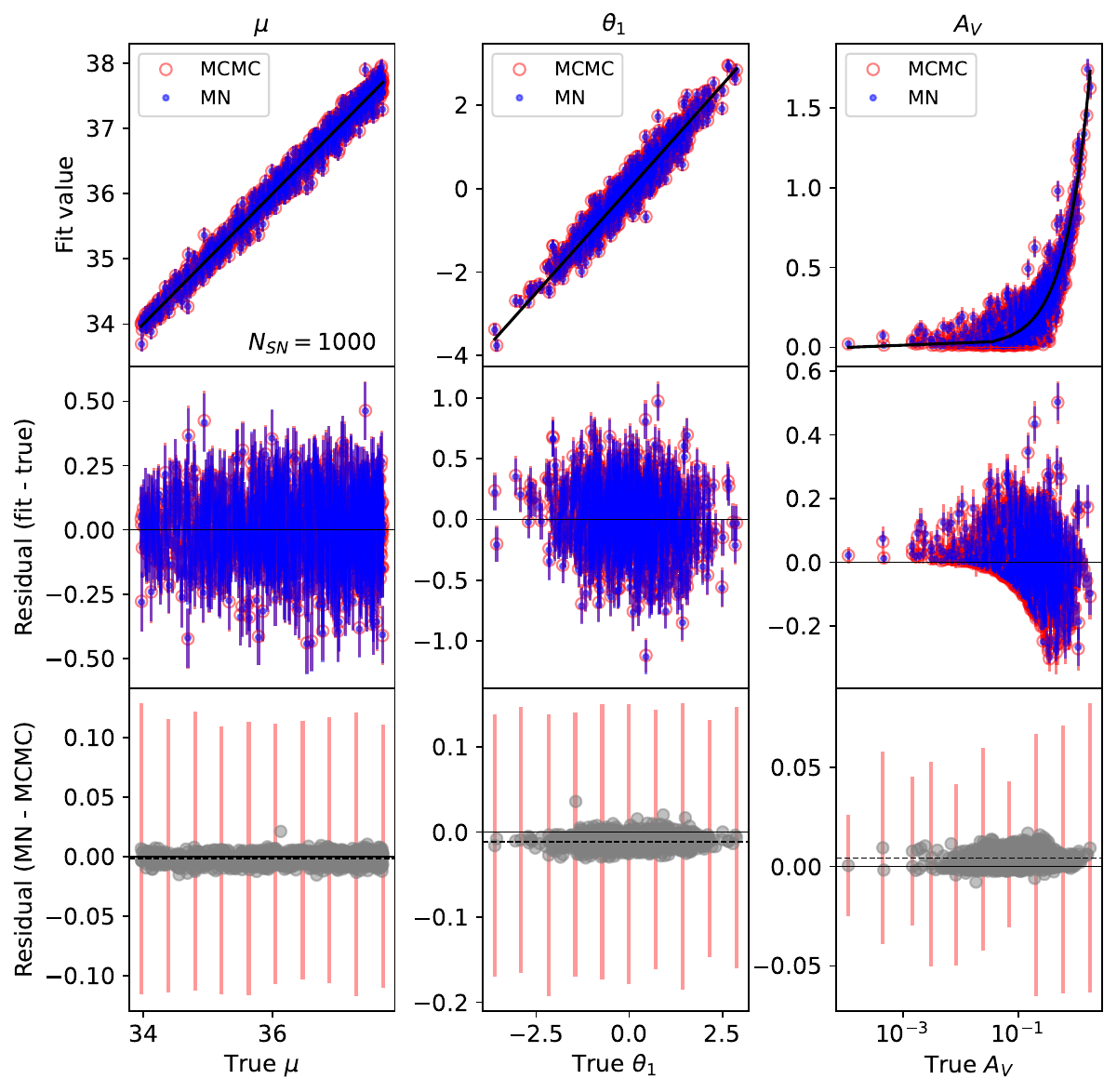} 
        \caption{A comparison of multivariate normal (MN) VI posterior fits (blue) and MCMC posterior fits (red) with the true parameter values for 1000 simulated SN Ia for $\mu$, $\theta_1$, and \av. The top row compares the median posterior value vs. the true parameter value, the middle row shows the residuals (fit - true) for MCMC and multivariate normal VI, and the bottom row the residuals (VI - MCMC) as a function of the true parameter value. Black dotted lines in the bottom row denote the median residual (MN - MCMC) value. The red error bars in the bottom row indicate the typical posterior standard deviation derived from MCMC fits to SNe at a set of 10 representative values spanning each parameter range. Note that the true \av{} is plotted on a $\log_{10}$ scale to better show the behavior at low \av.}
        \label{fig:multinormal_sims}
\end{figure*}

\section{Results and Validation}
\label{results_section}

Both variational approximations (MVZLTN and multivariate normal) and the Laplace Approximation are iteratively optimized, where each iteration consists of a single loss calculation and optimizer update step. In all cases, we fit the Laplace Approximation first for 15,000 iterations, which in practice was found to be sufficient for finding the MAP estimate. For the MVZLTN and multivariate normal guides, we then initialize on the median values of $\Delta t_\text{max}, \theta_1, A_V$, and $\mu$ from the Laplace approximation and train further for 10,000 iterations with the new guide. We use the \textsc{Adam} optimizer \citep{adam_optimizer} with a learning rate of 0.01 for the Laplace approximation and 0.005 for the MVZLTN and multivariate normal guides to optimize the ELBO computed with 5 ``particles'', indicating the number of samples from the surrogate posterior used to approximate the expectations in the ELBO (Eq. \ref{ELBO}). After optimizing the variational parameters of the guide for a given supernova, we then draw 1,000 parameter samples from the resulting approximate posterior distribution.

As a point of comparison, we also fit the same data with MCMC using the No-U-Turn Sampling ({\sc NUTS}) method in \textsc{NumPyro} \citep{nuts}. We use four chains with 250 warm-up samples and 250 additional samples in each, as is done by previous \textsc{BayeSN} analyses \citep{thorp21, thorp22, Grayling2024} to ensure a sufficient effective sample size (computed by \textsc{NumPyro} based on auto-correlation length of the chains; see \citealp{geyer92, geyer11, nuts, gelman_book}). Post-warmup, the NUTS sampler obtains effective samples at a rate $\approx0.5$--2 per iteration. Only the 250 post-warmup samples in each chain are used, yielding 1,000 total posterior samples (with effective sample sizes $\approx400$--2,000)\footnote{Effective sample sizes larger than the number of posterior samples (i.e.\ super-efficiency) are possible for NUTS, which can generate chains with negative auto-correlation (see discussion in \citealp{stan, vehtari21}).} for each SN.

\subsection{Results on Simulated Data}
\label{simulated_data_section}
Figures~\ref{fig:zltn_sims} and~\ref{fig:multinormal_sims} show the results of fitting 1000 simulated supernovae with MVZLTN VI and multivariate normal VI compared to the MCMC fits and the true parameter values used to generate the simulated light curves. The top row compares both VI and MCMC fits against the truth, the middle row shows the residuals of both VI and MCMC compared to the truth, and the bottom row shows the residuals between VI and MCMC.

Both the VI and MCMC posterior distributions are able to recover the true parameter values well. The median VI-true residuals (seen in the second row of Figures~\ref{fig:zltn_sims} and~\ref{fig:multinormal_sims}) standardized by the corresponding VI standard deviations for $\mu$, $\theta_1$, and $A_V$, respectively are -0.04$\sigma$, 0.04$\sigma$, and 0.11$\sigma$ for the MVZLTN posterior and -0.07$\sigma$, -0.05$\sigma$, and 0.15$\sigma$ for the multivariate normal posterior. For the MCMC approximation, the corresponding median standardized residuals are -0.04$\sigma$, 0.03$\sigma$, and 0.03$\sigma$ for $\mu$, $\theta_1$, and $A_V$, respectively. 

The residuals between the VI and MCMC parameter samples are small compared to the typical errors from the MCMC. The bottom row of Figures~\ref{fig:zltn_sims} and ~\ref{fig:multinormal_sims} shows the distribution of residual values (VI - MCMC) in the context of representative standard deviations of the MCMC marginal distributions for each of $\mu$, $\theta_1$, and \av. The median VI - MCMC residuals scaled by the MCMC standard deviations for $\mu$, $\theta_1$, and $A_V$ are 0.002$\sigma$, 0.02$\sigma$, and 0.06$\sigma$ for the MVZLTN posterior and -0.01$\sigma$, -0.07$\sigma$, and 0.07$\sigma$ for the multivariate normal posterior.

At $A_V \lesssim 0.01$, the VI residuals are almost exclusively positive relative to the truth. This is to be expected as we are constraining the marginal distribution of \av{} to be positive, hence for low values of true \av{} it is unlikely that the VI fit could underestimate them without being negative. At larger value of true $A_V$, there is a more even distribution of positive and negative residual values.

For \textsc{Variational-BayeSN}, the best-case scenario is that the VI approximation matches the performance of the MCMC, which would indicate no tradeoff between computational efficiency and the quality of the posterior approximation. However, this would also require the chosen form of the surrogate posterior to perfectly match that of the true posterior, which in reality is unlikely to be completely true. Thus, some variance is to be expected both between the MCMC samples and true parameters, and the VI approximation and MCMC samples.

As shown in the last row of Figures~\ref{fig:zltn_sims} and~\ref{fig:multinormal_sims} and the numerical summaries presented above, both the MVZLTN and multivariate normal VI approximations slightly overestimate the marginal values of $A_V$ compared to MCMC, while the multivariate normal VI also appears to underestimate $\theta_1$. Meanwhile, in both cases, there are neglible differences between the VI and MCMC estimates of the distance moduli $\mu$. The marginal distributions of the VI approximations will be further validated in Section~\ref{section:vsbc}.

\subsection{Results on Real Data}
\label{real_data_section}
\begin{figure}
    \centering
        \includegraphics[width=\columnwidth]{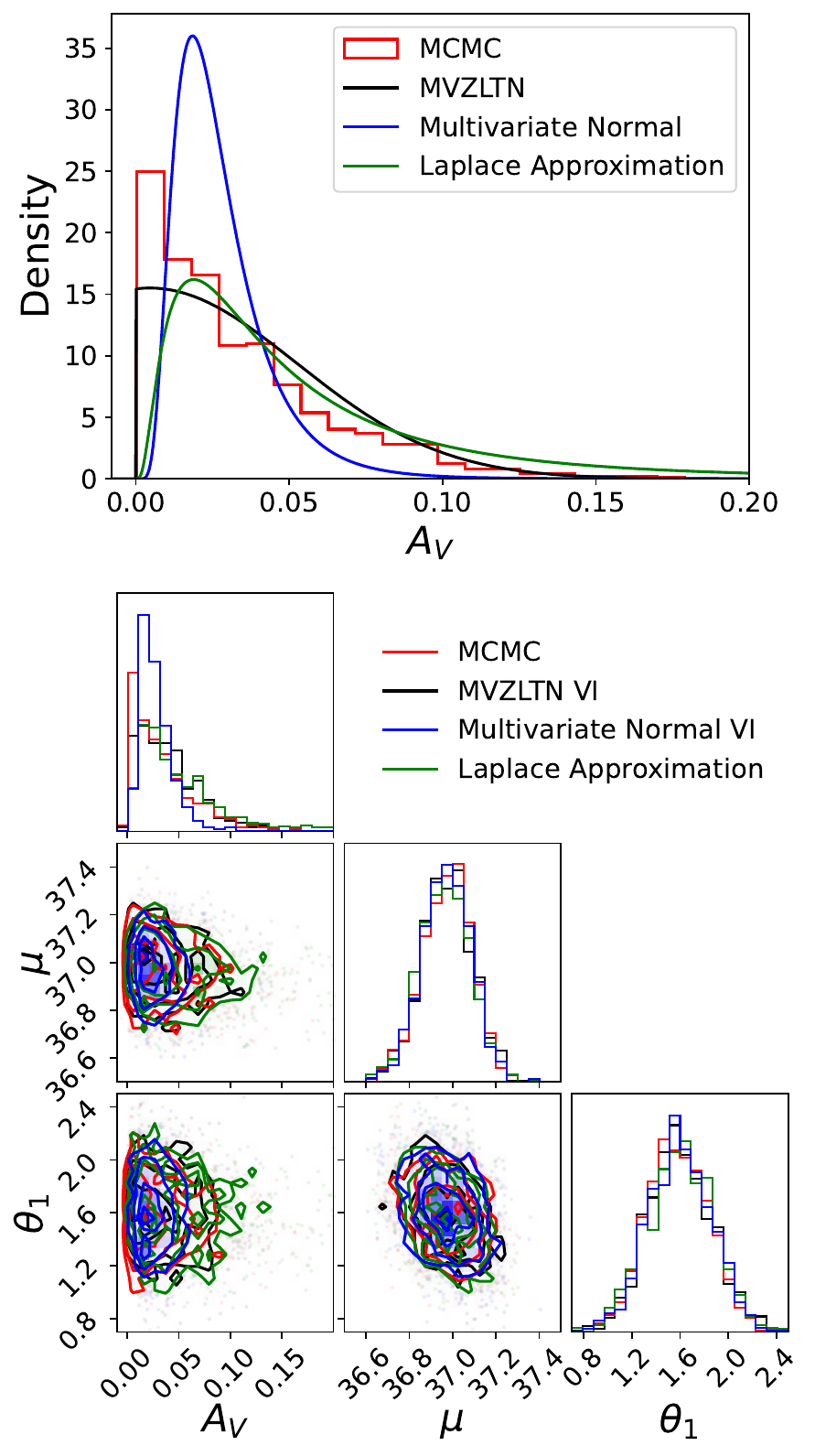} 
        \caption{(top) Zoomed-in plot of the estimated marginal posterior distributions over $A_V$. For the VI and Laplace approximation, we plot the analytic form of the approximate posterior (see Appendix \ref{zltn_marginals} for the MVZLTN marginals). (bottom) Corner plot showing the posterior distributions from MCMC (red), MVZLTN VI (black), multivariate normal VI (blue),  and Laplace Approximation (green), for SN2017cjv, an SN with low $A_V$. }
        \label{fig:foundation_single_low}
\end{figure}

\begin{figure}
    \centering
        \includegraphics[width=\columnwidth]{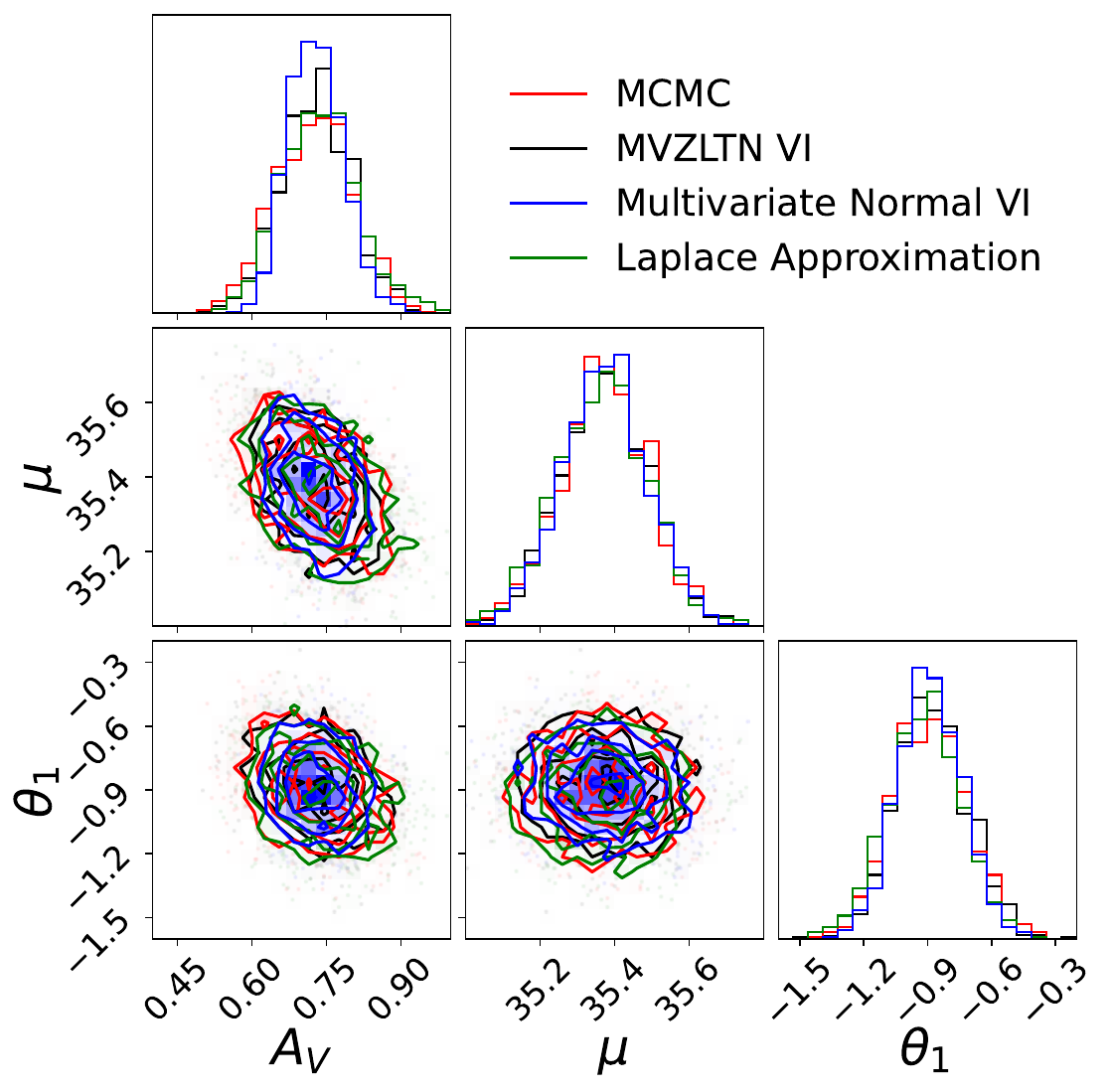}
        \caption{Corner plot showing the posterior distributions from MCMC (red), MVZLTN VI (black), multivariate normal VI (blue),  and Laplace Approximation (green), for ASASSN-16ay, a SN with high $A_V$.}
        \label{fig:foundation_single_high}
\end{figure}

We fit the Foundation Dataset (described in Section~\ref{foundation_section}), which consists of 157 real SNe Ia. Figure~\ref{fig:foundation_single_low} shows the marginal posterior for $A_V$, $\theta_1$, and $\mu$ for SN2017cjv, a low extinction SN Ia, with MCMC, MVZLTN VI, multivariate normal VI, and the Laplace approximation, while Figure~\ref{fig:foundation_single_high} shows the same for ASASSN-16ay, a highly extinguished SN Ia. For $A_V$ far from zero as in Figure~\ref{fig:foundation_single_high}, the four different types of posterior approximations agree quite well and are virtually indistinguishable. 

The top left subplot in Figure~\ref{fig:foundation_single_low} shows the marginal distribution of \av{} for each of the four posterior approximations. As expected the multivariate normal VI and the Laplace approximation show log-normal behavior with zero density as $A_V$ approaches 0. The MCMC distribution shows non-negligible density as $A_V$ approaches 0, which is most closely matched by the MVZLTN VI posterior. Of the variational approximations, the MVZLTN appears to model a low-extinction SN's posterior the best.

\begin{figure}
    \centering
        \includegraphics[width=\columnwidth]{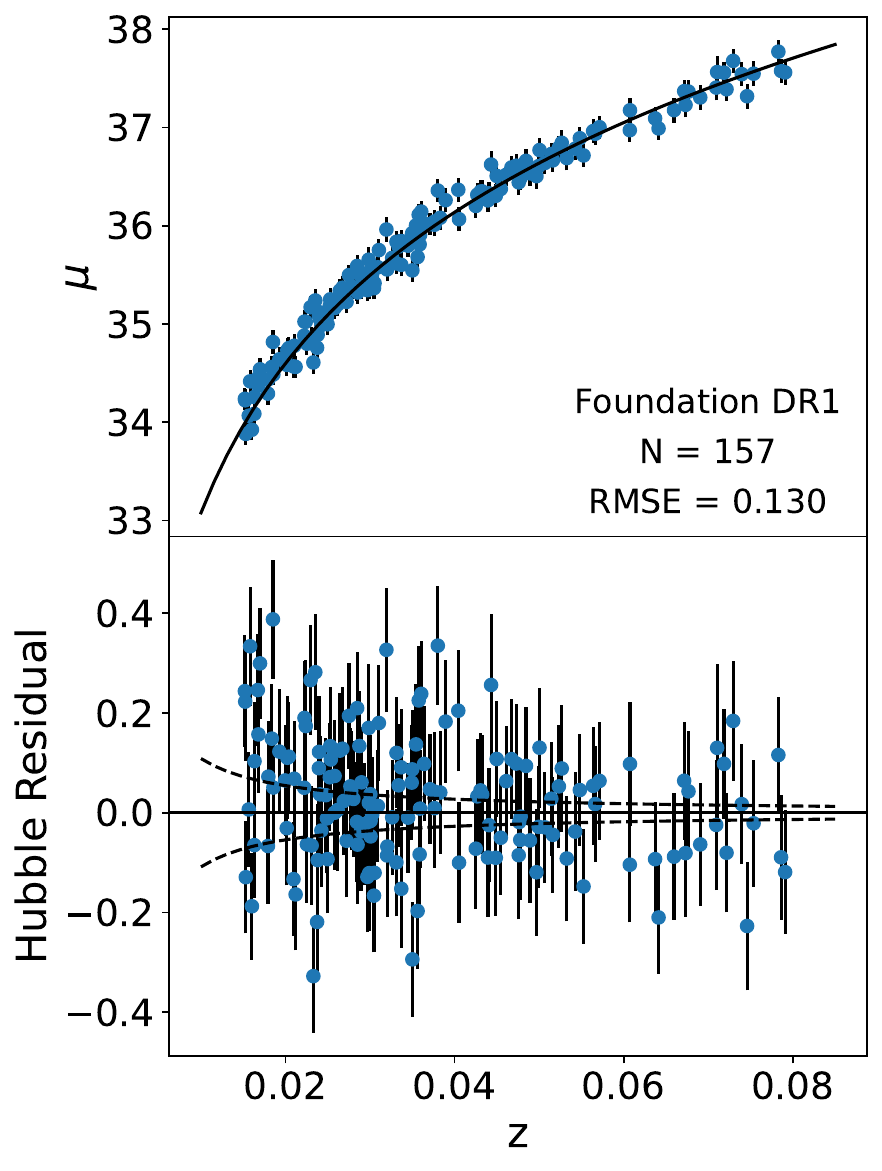} 
        \caption{(top) Hubble diagram of distance modulus $\mu$ vs. redshift $z$ for 157 supernovae in the Foundation dataset fit with MVZLTN VI. Black line indicates expected Hubble relation under flat $\Lambda$CDM with $H_0 = 73.24~\text{km}\,\text{s}^{-1}\,\text{Mpc}^{-1}$ and $\Omega_0 = 0.28$ \citep{riess2016}. Error bars indicate $1\sigma$ for each supernova. The RMSE for this dataset is 0.13. (bottom) Hubble residuals (difference between fitted $\mu$ values and expected Hubble relation shown in top panel) for 157 supernovae in Foundation dataset. Black horizontal line at zero is for reference, while dotted black curves represent the peculiar velocity uncertainty envelope. Error bars represent $1\sigma$.}
        \label{fig:hubble_diagram}
\end{figure}

Figure~\ref{fig:hubble_diagram} displays a Hubble diagram of all SNe in the Foundation dataset fit with MVZLTN VI. The root-mean squared error (RMSE) is 0.130~mag, which is consistent with that from the same analysis done by \citet{thorp21} using HMC. We perform the same analysis with the multivariate normal VI, Laplace Approximation and MCMC, which yield RMSE values of 0.129, 0.127, and 0.130~mag, respectively. 

\subsection{Validation of VI Performance}

It is possible to optimize the parameters for a surrogate posterior of a given form using VI but still not approximate the true posterior well. We use the methods for validating the performance of a VI model presented by \citet{yesbutdiditwork} (hereafter \citetalias{yesbutdiditwork}) to evaluate the quality of our variational approximations.

\subsubsection{Pareto-Smoothed Importance Sampling (PSIS)}

\begin{figure}
    \centering
        \includegraphics[width=\columnwidth]{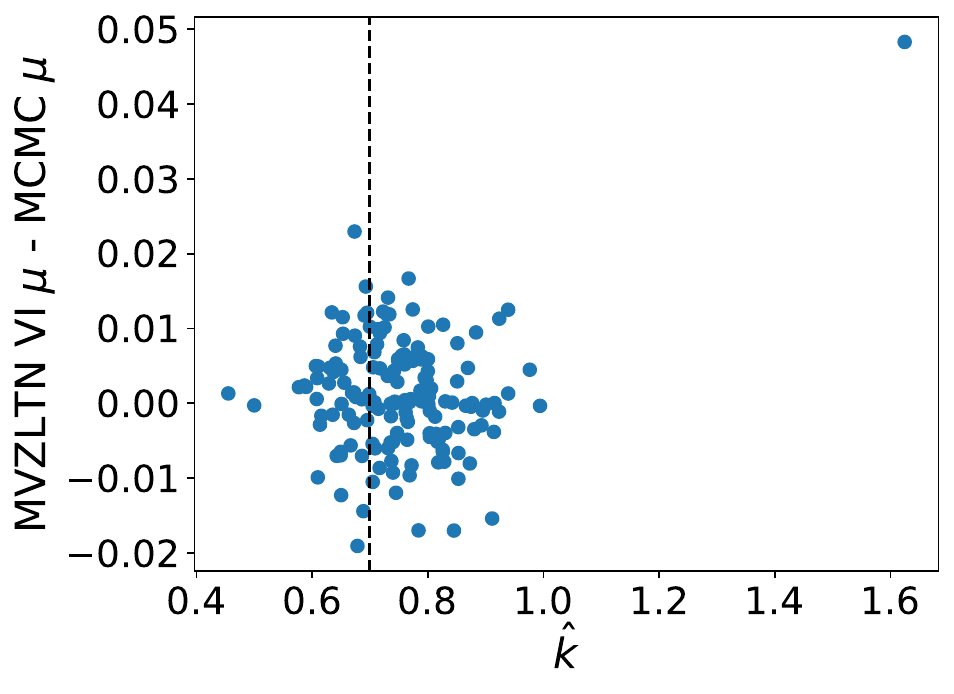}
        \caption{Median distance modulus $\mu$ residual (MVZLTN VI $-$ MCMC) vs.\ PSIS $\hat{k}$ diagnostic calculated from $N = 5000$ parameter samples for the Foundation dataset. Dotted line indicates $\hat{k} = 0.7$, a threshold for a good variational approximation of the posterior \citep{vehtari15}.}
        \label{fig:psis}
\end{figure}

The Pareto-Smoothed Importance Sampling (PSIS; \citealp{vehtari15}) diagnostic evaluates the goodness of fit of the surrogate posterior distribution compared to the true joint probability  (\citetalias{yesbutdiditwork}). For each of $N$ posterior samples $\bm{\phi}_{1},\dots,\bm{\phi}_N$, the importance ratio $r_n$ is calculated as:
\begin{equation}
    r_n = \frac{p(\bm{\phi}_n, x)}{q_\zeta(\bm{\phi}_n)}
\end{equation}
for data $x$, parameters $\bm{\phi}_n$, joint density $p(\bm{\phi}_n, x)$, and surrogate posterior $q_\zeta(\bm{\phi}_n)$. The $M$ largest importance ratios -- where $M = \min(N/5, 3 \sqrt{N}$) -- are fit to a generalized Pareto distribution yielding a shape parameter $\hat{k}$ \citep{vehtari15}. If $\hat{k} < 0.5$, this indicates that the variational approximation is a good approximation to the true joint density, while if $0.5 < \hat{k} < 0.7$, the approximation is still acceptable \citepalias{yesbutdiditwork}. A $\hat{k} > 0.7$ suggests that the variational approximation has large uncertainty \citepalias{yesbutdiditwork}. 

In calculating the PSIS diagnostic for our MVZLTN VI fits of the Foundation dataset, we fit the Foundation dataset as described at the beginning of Section~\ref{results_section} and use $N = 5000$ parameter samples from the posterior (using > 2000 samples as suggested by \citealp{vehtari15}) to calculate $\hat{k}$. Figure~\ref{fig:psis} shows the distribution of $\hat{k}$ values as a function of the difference in the median distance modulus $\mu$ from the MVZLTN VI and MCMC fits. The values of the residuals do not seem to correlate with the $\hat{k}$ diagnostic and, save for one outlier, appear approximately normally distributed around a \mutext residual of zero. \citet{vehtari15} note that performance of the PSIS diagnostic deteriorates as the number of parameters in the posterior increases, and \citet{huggins20} find that the diagnostic does not always correlate with the quality of the posterior approximation as intended. Overall, we do not find the PSIS $\hat{k}$ diagnostic to be effective in assessing the quality of our variational approximations, at least in terms of producing a distance modulus point estimate; however, we acknowledge the importance of validating the performance of our VI implementation.

\subsubsection{Variational Simulation-Based Calibration (VSBC)}
\label{section:vsbc}

The Variational Simulation-Based Calibration (VSBC) diagnostics determine whether a point estimate of a single parameter from its marginal distribution from a variational approximation is likely to be biased \citepalias{yesbutdiditwork}.  

For our simulated results shown in Section~\ref{simulated_data_section}, we generate 1,000 true parameter values from the priors shown in Section~\ref{data_section}, and simulate a set of SN Ia $griz$ light curves using each set of these parameters, corresponding to a single simulated SN Ia. We then fit the posterior using each of the outlined methods, and generate 1,000 samples from the posterior. To compute the VSBC diagnostics, we calculate $p_{s\phi}$, the probability of a posterior sample of model parameter $\phi$ from the variational approximation for simulated SN Ia $s$ being greater than the true value of $\phi$ (drawn from the priors listed in Section~\ref{data_section}) that was used in creating the simulated data for SN Ia $s$. In practice, this is simply the fraction of our 1,000 posterior samples that are below the true simulated parameter value:

\begin{equation}
    p_{s\phi} = \frac{\text{\# (posterior samples > true value)}}{\text{\# posterior samples}}
\end{equation}
for simulated SN Ia $s = {1,\dots,1000}$ and model parameters $\phi \in \{\mu, \theta_1, A_V\}$. We then investigate whether the distributions of these $p_{s\phi}$ values are symmetric across all simulated SNe for each model parameter $\phi$.  \citet{yesbutdiditwork} note that if the distribution is not symmetric, then the point estimate derived from the VI approximation is unlikely to be close to the median value of the true posterior.

\begin{figure*}
    \centering
        \includegraphics[width=\textwidth]{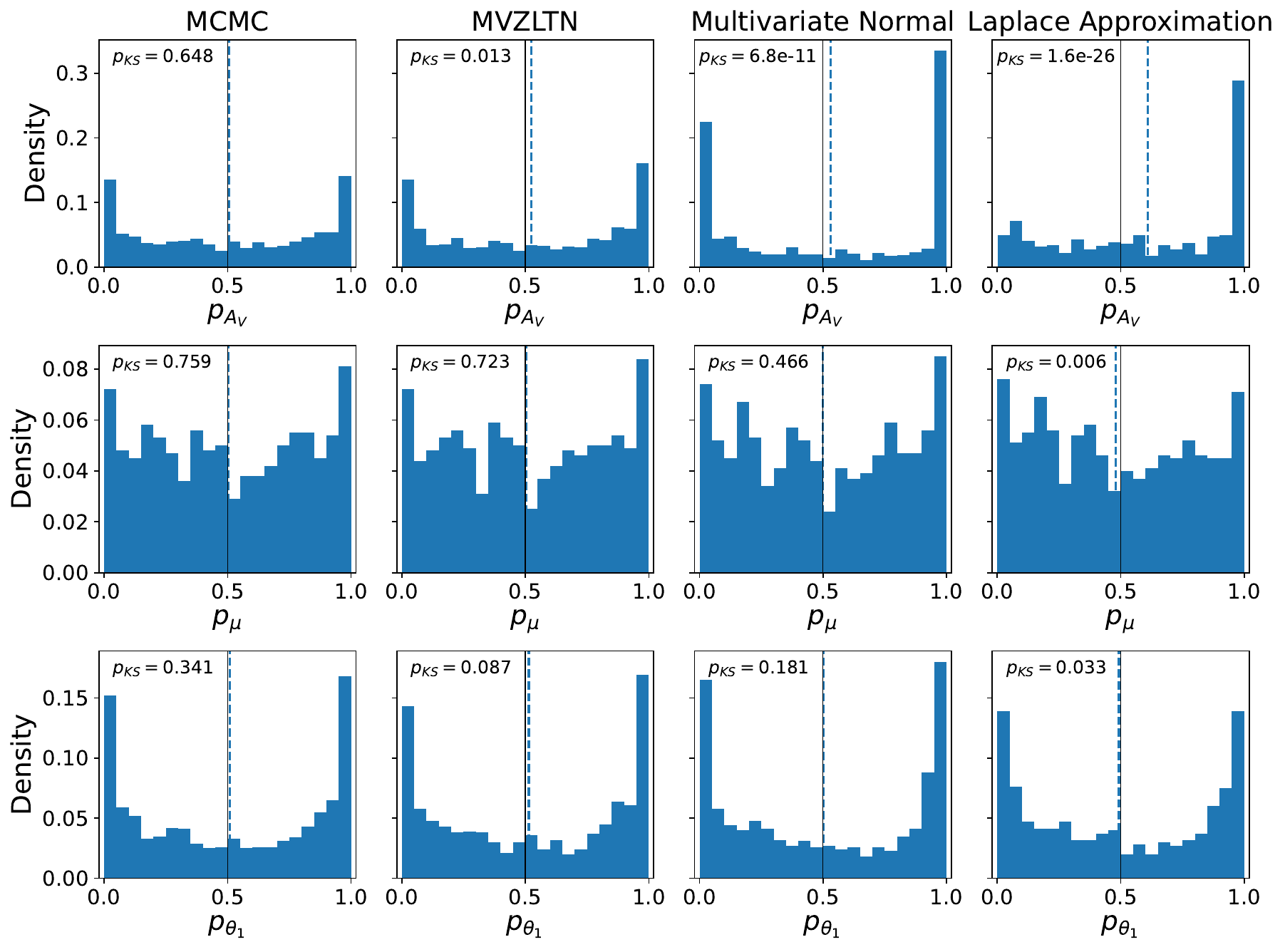}
        \caption{Results from VSBC diagnostic for MCMC, MVZLTN, multivariate normal, and Laplace Approximation posteriors, showing histograms of the probability of a single parameter sample from the posterior being less than the true parameter value, for 1000 simulated SNe. Dotted vertical line indicates mean probability, and solid black line indicates $p=0.5$ for comparison. }
        \label{fig:vsbc_results}
\end{figure*}

In each panel of Figure~\ref{fig:vsbc_results}, we show the histogram of $p_{s\phi}$ for $\phi = \mu$, \av, or $\theta_1$ across all supernovae $s=1\dots1000$ for each type of posterior fit. To investigate the symmetry of each marginal distribution, we perform a 2-sided Kolmogorov--Smirnov (KS) test \citep{kolmogorov33, smirnov48, massey1951kolmogorov} on the distributions of $p_{\phi}$ and $1 - p_{\phi}$ as in \citetalias{yesbutdiditwork}. The KS test yields a frequentist $p$-value, here denotes as $p_{KS}$, which if less than 0.05 suggests that $p_{\phi}$ and $1-p_{\phi}$ are unlikely to be drawn from the same distribution and thus the distribution of $p_{\phi}$ is not symmetric. These $p_{KS}$-values are annotated on each subplot in Figure~\ref{fig:vsbc_results}. Note that this technically violates the requirement of the KS test that the two distributions being compared are independent; however, we chose to follow the procedure for the VSBC diagnostic as described by \citetalias{yesbutdiditwork}.

A given marginal distribution from a variational approximation that closely resembles the true posterior would be symmetric about $p_\phi = 0.5$, denoted by a black vertical line in Figure~\ref{fig:vsbc_results}. The results of the VSBC diagnostics show that there is a systematic offset in the point estimates from the Laplace approximation for all three parameters. While the two VI methods tend to overestimate $A_V$, the marginal distributions of $\theta_1$ and \mutext remain centered around the true values. The VSBC diagnostics find that the MCMC marginal distributions are unbiased, suggesting both VI approximations slightly overestimate $A_V$ when compared to the MCMC, but these offsets are small compared to the MCMC uncertainties, as seen in the bottom row of Figures 1 and 2 and the numerical summaries discussed in Section \ref{simulated_data_section}.

 Whereas the PSIS diagnostic evaluates performance on the 28-dimensional joint posterior, the VSBC diagnostics solely operate on the marginal distribution of one parameter, ignoring other parameters even though they may covary. In practice, the most common application of SNe Ia SED fitting models such as \bayesn{} is to fit the distance-redshift relation for cosmological parameter estimation. For this, the posterior point estimate of the photometric distance modulus $\mu$ marginalized over all other light curve parameters is needed, and Figure~\ref{fig:vsbc_results} shows that these point estimates are unbiased in any of the variational approximations.

\subsection{Computational Performance}
The runtime of our \numpyro{} code depends on how many chains and samples are run for MCMC and how many iterations used in VI. We used the {\tt jax.vmap()} method to vectorize the code to fit multiple SNe in parallel \citep{jax}. The implementation of \bayesn{} in \numpyro{} by \citet{Grayling2024} enables the use of GPUs for further computational speedup.

Table~\ref{tab:runtime_table} shows the total and per-SN runtimes for fitting the Foundation Dataset, which consists of 157 SNe Ia, and a simulated population of 1000 SNe on both an Apple M2 Pro CPU and an NVIDIA Ampere A100 GPU. On CPU, MCMC was run in parallel across four cores (with one chain per core). In GPU mode, MCMC was run in parallel across four separate GPUs. In contrast, VI required only a single CPU core or GPU. The runtimes in Table~\ref{tab:runtime_table} show that approximating the posterior with VI is about an order of magnitude faster in addition using less computational resources.

Note that while the per-SN time is listed for comparison between methods, light curves are fit in parallel and the listed time is not representative of the time it would take to fit a single supernova by itself. For the GPUs in particular, the per-SN runtime decreases when fitting larger SN populations. When fitting few SNe, it is likely faster to use the CPU. Each SN was fit as described at the beginning of Section~\ref{results_section}.

\begin{table*}
    \centering
    \begin{threeparttable}
        \caption{A table of runtimes (wall clock time) for fitting the Foundation dataset (157 SNe) and a simulated population (1000 SNe). Per SN runtimes are illustrative\tnote{*}.}
        \label{tab:runtime_table}
        \begin{tabular}{c c c c c c c c c}
            \toprule
            Method & \multicolumn{4}{c}{CPU\tnote{a}} & \multicolumn{4}{c}{GPU\tnote{b}} \\
            \cmidrule(lr){2-5} \cmidrule(l){6-9}
            & \multicolumn{2}{c}{Foundation} & \multicolumn{2}{c}{Simulations} & \multicolumn{2}{c}{Foundation} & \multicolumn{2}{c}{Simulations} \\
            \cmidrule(lr){2-3} \cmidrule(lr){4-5} \cmidrule(lr){6-7} \cmidrule(l){8-9}
            & total & per SN & total & per SN & total & per SN & total & per SN\\
            \midrule
            MCMC\tnote{\P}  & 3h\,48m & 1m\,27s & 16h\,24m & 1m & 4m 42s & 2s & 11m 20s & 0.7s\\
            Laplace Approx.\tnote{\S} & 7m 15s & 3s & 38m 47s & 2s & 43s & 0.3s & 1m 15s & 0.1s\\ 
            MultiNormal VI\tnote{\S} & 26m 37s & 10s & 2h\,24m & 9s & 42s & 0.3s & 3m 2s & 0.2s\\ 
            MVZLTN VI\tnote{\S} & 28m 22s & 11s & 2h\,43m & 10s & 54s & 0.3s & 2m 45s & 0.2s\\ 
            \bottomrule
        \end{tabular}
        \begin{tablenotes}
            \item [a] CPU runtimes are based on an Apple M2 Pro
            \item [b] GPU runtimes are based on an NVIDIA Ampere A100
            \item [\P] MCMC CPU/GPU results used four parallel chains, with one core/GPU per chain
            \item[\S] VI / Laplace CPU results were run using a single core
            \item[*] On GPU, the fractional speedup is greater when fitting large batches of SNe. All SNe were fit in parallel, which leads to a considerable performance increase on GPU. However, this is not necessarily the case when running on CPU. For a like-for-like comparison between all methods, we ran parallel fits on both CPU and GPU for this work. For a small sample of objects being fit on CPU, faster per-SN times can be achieved by fitting objects in series, which the public release of \bayesn{} supports.
        \end{tablenotes}
    \end{threeparttable}
\end{table*}

\section{Discussion}
\label{discussion_section}
We use both simulated and real data to evaluate the performance of our variational approximations. While the real-world use case of these methods is inferring parameters from observations with no known true values, fitting simulated data where the true parameters are known allows for additional insight into the accuracy and limitations of both VI and MCMC. In practice, a performant VI approximation will yield similar results to MCMC, which our results show on both simulated and real data.

There are several reasons to use VI over MCMC. The first is increased computational efficiency in both time and computing resources. The variational approximation also provides an analytic expression for the posterior distribution, which enables several types of additional analyses, such as calculating importance sampling ratios or the Hessian matrix. An analytic expression for the posterior also enables robust uncertainty propagation, and can be used as a prior in downstream analyses. MCMC, nested sampling, or other posterior sampling methods are likely still the best option for models where the form of the posterior cannot be assumed, is multimodal, or is another form that is difficult to express analytically. However, the compressed representation of the posterior in terms of the variational parameters $\bm{\zeta}$ is also advantageous in terms of data volume, as saving many posterior samples is infeasible in the large-data regime (see e.g.\ discussion in \citealp{malz18}).

Here we consider a particularly difficult use case with one parameter that is constrained to be nonnegative. The performance of all of the different types of variational posteriors on the non-constrained parameters is quite good, and for most use cases using a Multivariate Gaussian guide or even the Laplace approximation will achieve similar results to MCMC in a fraction of the computational time (as seen in Figure~\ref{fig:foundation_single_high}) while also providing an analytic posterior and robust uncertainties. This is especially promising in light of the expected quantity of data from the Vera C. Rubin Observatory/LSST in the coming years. 

The ZLTN distribution is asymmetric, and so, when using it to approximate the marginal distribution of $A_V$, the mode and the median are not the same. However, we found in practice that using the median sample as the point estimate led to better results. In calculating point estimates of $\mu$, we marginalize over the other parameters. While it is most important to estimate the distance modulus \mutext accurately for precision cosmology, this cannot happen without proper modeling of $A_V$. There is a posterior covariance between \mutext and $A_V$ as both have a dimming effect. The degeneracy is not perfect, as $A_V$ has a wavelength-dependent effect. Nevertheless, probabilistically modeling the impact of dust on the SN Ia SED is critical to accurate distance estimation.

KL divergence is not symmetric ($KL[P || Q] \neq KL[Q || P]$), and this distinction can have important implications \citep{margossian2024}. VI uses the ELBO as its loss function, which minimizes the reverse KL divergence $D_\text{KL}[q_\zeta(\phi) || p(\phi, x)]$ between the true posterior $p(\phi, x)$ and the surrogate posterior $q_\zeta(\phi)$. When approximating a multimodal distribution with a normal distribution, the reverse KL divergence is ``mode-locking'', meaning that it will perfectly fit one mode of the distribution with zero density around the other modes \citep{minka2005divergence, turner2011two}. By contrast, the forward KL divergence $D_\text{KL}[p(\phi, x) || q_\zeta(\phi)]$ is ``mode-averaging'', meaning that it will approximate a multimodal distribution by spanning its density across all modes and any areas in between modes \citep{minka2005divergence, turner2011two}. There are other algorithms, such as expectation propagation, that use the forward KL divergence to avoid getting ``stuck'' in one mode of the posterior \citep{minka2013expectation}. In this work, we are not approximating a multimodal posterior distribution, but this is an important consideration for generalized use of VI \citep{margossian2024}. 

\section{Future Work \& Conclusions}
\label{conclusions}

In this work we present a VI implementation of \bayesn{} using \numpyro{} with multivariate normal and MVZLTN surrogate posterior and compare the results on both simulated and real data to MCMC and Laplace Approximation posteriors. We evaluate the PSIS and VSBC diagnostic to validate the quality of our variational approximation, and show that VI can achieve similar results to MCMC but with reduced runtime and computational resources.

All of the code for this work is publicly available at \url{https://github.com/asmuzsoy/bayesn-VI-paper}. The code for the custom {\tt AutoGuide} used to fit the MVZLTN surrogate posterior can easily be modified to allow other types of custom posterior forms for a single variable while still maintaining a fully conditional multivariate distribution. The VI implementation has also been incorporated into the larger public \bayesn{} release by \citet{Grayling2024}, available at \url{https://github.com/bayesn/bayesn}.

A promising future avenue would be to extend the application of VI to the \textsc{BayeSN} training process, where the model's hyperparameters are inferred. The work we have done here suggests that the possibility for speed-up is large when switching to VI over MCMC. An additional benefit would be that analytic approximate posteriors on the hyperparameters could be easily propagated into downstream analyses. A further extension that would be valuable in the LSST is the inclusion of redshift as a free parameter, so uncertainties associated with photometric host-galaxy redshifts can be marginalized over. This could easily be included in the existing \textsc{BayeSN} framework. Since photometric redshift PDFs (which would be used as a prior in a \textsc{BayeSN} light curve fit) are typically non-analytic and non-Gaussian, they likely be provided in a summarized form such as quantiles \citep{malz18}. Frameworks have been developed for the inclusion of such priors in probabilistic programming languages \citep[e.g.][]{perepolkin23}\footnote{See also the earlier approach by B.\ Goodrich based on Chebyshev polynomials: \url{https://github.com/bgoodri/StanCon2020}. Preliminary investigations using the Stan implementation of \textsc{BayeSN} showed this to be viable in an MCMC analysis (S.~M.\ Ward, private communication). This will be integrated into a future release of the NumPyro \textsc{BayeSN} code: \url{https://github.com/bayesn/bayesn/pull/37}.}.

There are also alternate methods to potentially better model the dust extinction coefficient $A_V$. While we model its posterior with a ZLTN distribution with all other variables normal conditional on $A_V$, the structure of the joint posterior could also be modeled with a copula for a greater deal of flexibility \citep{tran15, han16}. Within astronomy, copulas have recently been used by \citet{patil23} to model the distribution of stellar chemical abundances in the Milky Way. Modeling the effect of dust extinction is critical to accurate estimation of distances to SN Ia for precision cosmology. 

There have been multiple examples of applying Variational Auto-Encoders (VAEs), which use neural networks to learn nonlinear relationships, to extract physical parameters from transient light curves \citep{parsnip, villar_vae_classifier}. \bayesn{} explicitly models physical parameters such as the distance modulus \mutext and dust extinction parameter $A_V$, and includes additional parameters to model the remaining variance in the SED (see Equation~\ref{eq:bayesnsed}). A variation on \bayesn{} could combine these two approaches, explicitly modeling the SED dependence on physical parameters while allowing a neural network to learn a nonlinear representation of the rest of the model. This would not be as interpretable as the current \bayesn{} model, but would leverage the nonlinear representation power of neural networks while explicitly maintaining the known key physical parameters.

As the size of astronomical datasets increase and in anticipation of the Rubin Observatory/LSST, variational approximations such as the one in this work could be key to accelerating data analysis pipelines that are currently using slower sampling methods. In the era of precision cosmology, being able to use as much data as possible without draining expensive computational time and resources enables robust estimates and tighter constraints for cosmological parameters.

\section*{Acknowledgements}
ASMU was supported by a Churchill Scholarship and a National Science Foundation Graduate Research Fellowship. ST was supported by the Cambridge Centre for Doctoral Training in Data-Intensive Science funded by the UK Science and Technology Facilities Council (STFC), and in part by the European Research Council (ERC) under the European Union’s Horizon 2020 research and innovation programme (Grant Agreement No.\ 101018897 CosmicExplorer). MG and KSM acknowledge funding from the European Research Council under the European Union’s Horizon 2020 research and innovation programme (ERC Grant Agreement No. 101002652). This project has been made possible through the ASTROSTAT-II collaboration, enabled by the EU Horizon 2020, Marie Skłodowska-Curie Grant Agreement No. 873089. This manuscript is based on work that was originally developed in the MPhil thesis of ASMU at the University of Cambridge \citep{Uzsoy22}.

ASMU would like to thank Douglas Finkbeiner, Andrew Saydjari, Souhardya Sengupta, Tamara Broderick, Ryan Giordano, Gautham Narayan, Rick Kessler, Suhail Dhawan, Sam Ward, Ben Boyd, Erin Hayes, Ashley Villar, and the Villar Time-Domain Astrophysics group for helpful discussions and suggestions.
 

This work was performed using resources provided by the Cambridge Service for Data Driven Discovery (CSD3) operated by the University of Cambridge Research Computing Service (\url{www.csd3.cam.ac.uk}), provided by Dell EMC and Intel using Tier-2 funding from the Engineering and Physical Sciences Research Council (capital grant EP/T022159/1), and DiRAC funding from the Science and Technology Facilities Council (\url{www.dirac.ac.uk}).

This work made use of the {\tt pyro} \citep{pyro}, {\tt numpyro} \citep{numpyro}, {\tt corner} \citep{corner}, {\tt numpy} \citep{numpy}, {\tt scipy} \citep{2020SciPy-NMeth}, {\tt matplotlib} \citep{matplotlib}, {\tt astropy} \citep{astropy}, {\tt extinction} \citep{extinction}, {\tt arviz} \citep{arviz_2019} and {\tt jax} \citep{jax} Python packages.

\section*{Data Availability}


The data for the 180 supernovae in the Foundation DR1 cosmology sample \citep{foley18, jones19} are publicly available at \url{https://github.com/djones1040/Foundation_DR1}.

The simulated data shown, as well as all code used in this work, are publicly available on GitHub at \url{https://github.com/asmuzsoy/bayesn-VI-paper}.



\bibliographystyle{mnras}
\bibliography{main} 

\begin{thebibliography}{}
\makeatletter
\relax
\def\mn@urlcharsother{\let\do\@makeother \do\$\do\&\do\#\do\^\do\_\do\%\do\~}
\def\mn@doi{\begingroup\mn@urlcharsother \@ifnextchar [ {\mn@doi@} {\mn@doi@[]}}
\def\mn@doi@[#1]#2{\def\@tempa{#1}\ifx\@tempa\@empty \href {http://dx.doi.org/#2} {doi:#2}\else \href {http://dx.doi.org/#2} {#1}\fi \endgroup}
\def\mn@eprint#1#2{\mn@eprint@#1:#2::\@nil}
\def\mn@eprint@arXiv#1{\href {http://arxiv.org/abs/#1} {{\tt arXiv:#1}}}
\def\mn@eprint@dblp#1{\href {http://dblp.uni-trier.de/rec/bibtex/#1.xml} {dblp:#1}}
\def\mn@eprint@#1:#2:#3:#4\@nil{\def\@tempa {#1}\def\@tempb {#2}\def\@tempc {#3}\ifx \@tempc \@empty \let \@tempc \@tempb \let \@tempb \@tempa \fi \ifx \@tempb \@empty \def\@tempb {arXiv}\fi \@ifundefined {mn@eprint@\@tempb}{\@tempb:\@tempc}{\expandafter \expandafter \csname mn@eprint@\@tempb\endcsname \expandafter{\@tempc}}}

\bibitem[\protect\citeauthoryear{{Abbott} et~al.,}{{Abbott} et~al.}{2024}]{des24}
{Abbott} T.~M.~C.,  et~al., 2024, \mn@doi [\apjl] {10.3847/2041-8213/ad6f9f}, \href {https://ui.adsabs.harvard.edu/abs/2024ApJ...973L..14A} {973, L14}

\bibitem[\protect\citeauthoryear{{Aleo} et~al.,}{{Aleo} et~al.}{2023}]{aleo23}
{Aleo} P.~D.,  et~al., 2023, \mn@doi [\apjs] {10.3847/1538-4365/acbfba}, \href {https://ui.adsabs.harvard.edu/abs/2023ApJS..266....9A} {266, 9}

\bibitem[\protect\citeauthoryear{{Astier} et~al.,}{{Astier} et~al.}{2006}]{astier06}
{Astier} P.,  et~al., 2006, \mn@doi [\aap] {10.1051/0004-6361:20054185}, \href {https://ui.adsabs.harvard.edu/abs/2006A&A...447...31A} {447, 31}

\bibitem[\protect\citeauthoryear{{Astropy Collaboration} et~al.,}{{Astropy Collaboration} et~al.}{2013}]{astropy}
{Astropy Collaboration} et~al., 2013, \mn@doi [\aap] {10.1051/0004-6361/201322068}, \href {http://adsabs.harvard.edu/abs/2013A%26A...558A..33A} {558, A33}

\bibitem[\protect\citeauthoryear{{Barbary}}{{Barbary}}{2021}]{extinction}
{Barbary} K.,  2021, {extinction: Dust extinction laws}, Astrophysics Source Code Library, record ascl:2102.026

\bibitem[\protect\citeauthoryear{Bingham et~al.,}{Bingham et~al.}{2019}]{pyro}
Bingham E.,  et~al., 2019, J.\ Machine Learning Res., \href {https://ui.adsabs.harvard.edu/abs/2018arXiv181009538B} {20, 1}

\bibitem[\protect\citeauthoryear{Blei, Kucukelbir  \& McAuliffe}{Blei et~al.}{2017}]{VI_review}
Blei D.~M.,  Kucukelbir A.,   McAuliffe J.~D.,  2017, \mn@doi [J.\ American Statistical Association] {10.1080/01621459.2017.1285773}, \href {https://ui.adsabs.harvard.edu/abs/2016arXiv160100670B} {112, 859}

\bibitem[\protect\citeauthoryear{{Boone}}{{Boone}}{2021}]{parsnip}
{Boone} K.,  2021, \mn@doi [\aj] {10.3847/1538-3881/ac2a2d}, \href {https://ui.adsabs.harvard.edu/abs/2021AJ....162..275B} {162, 275}

\bibitem[\protect\citeauthoryear{Bradbury et~al.,}{Bradbury et~al.}{2018}]{jax}
Bradbury J.,  et~al., 2018, {JAX}: composable transformations of {P}ython+{N}um{P}y programs, \url {http://github.com/google/jax}

\bibitem[\protect\citeauthoryear{{Branch} \& {Tammann}}{{Branch} \& {Tammann}}{1992}]{branch92}
{Branch} D.,  {Tammann} G.~A.,  1992, \mn@doi [\araa] {10.1146/annurev.aa.30.090192.002043}, \href {https://ui.adsabs.harvard.edu/abs/1992ARA&A..30..359B} {30, 359}

\bibitem[\protect\citeauthoryear{{Brout} \& {Scolnic}}{{Brout} \& {Scolnic}}{2021}]{brout21}
{Brout} D.,  {Scolnic} D.,  2021, \mn@doi [\apj] {10.3847/1538-4357/abd69b}, \href {https://ui.adsabs.harvard.edu/abs/2021ApJ...909...26B} {909, 26}

\bibitem[\protect\citeauthoryear{{Brout} et~al.,}{{Brout} et~al.}{2019}]{brout19}
{Brout} D.,  et~al., 2019, \mn@doi [\apj] {10.3847/1538-4357/ab06c1}, \href {https://ui.adsabs.harvard.edu/abs/2019ApJ...874..106B} {874, 106}

\bibitem[\protect\citeauthoryear{{Burns} et~al.,}{{Burns} et~al.}{2011}]{burns11}
{Burns} C.~R.,  et~al., 2011, \mn@doi [\aj] {10.1088/0004-6256/141/1/19}, \href {https://ui.adsabs.harvard.edu/abs/2011AJ....141...19B} {141, 19}

\bibitem[\protect\citeauthoryear{{Burns} et~al.,}{{Burns} et~al.}{2014}]{burns14}
{Burns} C.~R.,  et~al., 2014, \mn@doi [\apj] {10.1088/0004-637X/789/1/32}, \href {https://ui.adsabs.harvard.edu/abs/2014ApJ...789...32B} {789, 32}

\bibitem[\protect\citeauthoryear{{Commins}}{{Commins}}{2004}]{commins04}
{Commins} E.~D.,  2004, \mn@doi [\nar] {10.1016/j.newar.2003.12.035}, \href {https://ui.adsabs.harvard.edu/abs/2004NewAR..48..567C} {48, 567}

\bibitem[\protect\citeauthoryear{{Conley} et~al.,}{{Conley} et~al.}{2008}]{SiFTO}
{Conley} A.,  et~al., 2008, \mn@doi [\apj] {10.1086/588518}, \href {https://ui.adsabs.harvard.edu/abs/2008ApJ...681..482C} {681, 482}

\bibitem[\protect\citeauthoryear{Cook, Gelman  \& Rubin}{Cook et~al.}{2006}]{cook06}
Cook S.~R.,  Gelman A.,   Rubin D.~B.,  2006, \mn@doi [J.\ Comput.\ Graphical Statistics] {10.1198/106186006X136976}, 15, 675

\bibitem[\protect\citeauthoryear{{\vphantom{DeSoto}de Soto} et~al.,}{{\vphantom{DeSoto}de Soto} et~al.}{2024}]{desoto24}
{\vphantom{DeSoto}de Soto} K.~M.,  et~al., 2024, \mn@doi [\apj] {10.3847/1538-4357/ad6a4f}, \href {https://ui.adsabs.harvard.edu/abs/2024ApJ...974..169D} {974, 169}

\bibitem[\protect\citeauthoryear{{Dhawan} et~al.,}{{Dhawan} et~al.}{2022}]{ztf1}
{Dhawan} S.,  et~al., 2022, \mn@doi [\mnras] {10.1093/mnras/stab3093}, \href {https://ui.adsabs.harvard.edu/abs/2022MNRAS.510.2228D} {510, 2228}

\bibitem[\protect\citeauthoryear{{Dhawan}, {Thorp}, {Mandel}, {Ward}, {Narayan}, {Jha}  \& {Chant}}{{Dhawan} et~al.}{2023}]{dhawan23}
{Dhawan} S.,  {Thorp} S.,  {Mandel} K.~S.,  {Ward} S.~M.,  {Narayan} G.,  {Jha} S.~W.,   {Chant} T.,  2023, \mn@doi [\mnras] {10.1093/mnras/stad1590}, \href {https://ui.adsabs.harvard.edu/abs/2023MNRAS.524..235D} {524, 235}

\bibitem[\protect\citeauthoryear{{Draine}}{{Draine}}{2003}]{dust_review}
{Draine} B.~T.,  2003, \mn@doi [\araa] {10.1146/annurev.astro.41.011802.094840}, \href {https://ui.adsabs.harvard.edu/abs/2003ARA&A..41..241D} {41, 241}

\bibitem[\protect\citeauthoryear{{Feroz}, {Hobson}  \& {Bridges}}{{Feroz} et~al.}{2009}]{multinest}
{Feroz} F.,  {Hobson} M.~P.,   {Bridges} M.,  2009, \mn@doi [\mnras] {10.1111/j.1365-2966.2009.14548.x}, \href {https://ui.adsabs.harvard.edu/abs/2009MNRAS.398.1601F} {398, 1601}

\bibitem[\protect\citeauthoryear{{Fitzpatrick}}{{Fitzpatrick}}{1999}]{fitzpatrick99}
{Fitzpatrick} E.~L.,  1999, \mn@doi [\pasp] {10.1086/316293}, \href {https://ui.adsabs.harvard.edu/abs/1999PASP..111...63F} {111, 63}

\bibitem[\protect\citeauthoryear{{Foley} et~al.,}{{Foley} et~al.}{2018}]{foley18}
{Foley} R.~J.,  et~al., 2018, \mn@doi [\mnras] {10.1093/mnras/stx3136}, \href {https://ui.adsabs.harvard.edu/abs/2018MNRAS.475..193F} {475, 193}

\bibitem[\protect\citeauthoryear{{Foreman-Mackey}}{{Foreman-Mackey}}{2016}]{corner}
{Foreman-Mackey} D.,  2016, \mn@doi [J.\ Open Source Software] {10.21105/joss.00024}, \href {https://ui.adsabs.harvard.edu/abs/2016JOSS....1...24F} {1, 24}

\bibitem[\protect\citeauthoryear{{Foreman-Mackey}, {Hogg}, {Lang}  \& {Goodman}}{{Foreman-Mackey} et~al.}{2013}]{emcee}
{Foreman-Mackey} D.,  {Hogg} D.~W.,  {Lang} D.,   {Goodman} J.,  2013, \mn@doi [\pasp] {10.1086/670067}, \href {https://ui.adsabs.harvard.edu/abs/2013PASP..125..306F} {125, 306}

\bibitem[\protect\citeauthoryear{{Freedman}}{{Freedman}}{2021}]{freedman2021}
{Freedman} W.~L.,  2021, \mn@doi [\apj] {10.3847/1538-4357/ac0e95}, \href {https://ui.adsabs.harvard.edu/abs/2021ApJ...919...16F} {919, 16}

\bibitem[\protect\citeauthoryear{{Friedman} et~al.,}{{Friedman} et~al.}{2015}]{friedman15}
{Friedman} A.~S.,  et~al., 2015, \mn@doi [\apjs] {10.1088/0067-0049/220/1/9}, \href {https://ui.adsabs.harvard.edu/abs/2015ApJS..220....9F} {220, 9}

\bibitem[\protect\citeauthoryear{{Frieman} et~al.,}{{Frieman} et~al.}{2008}]{frieman08}
{Frieman} J.~A.,  et~al., 2008, \mn@doi [\aj] {10.1088/0004-6256/135/1/338}, \href {https://ui.adsabs.harvard.edu/abs/2008AJ....135..338F} {135, 338}

\bibitem[\protect\citeauthoryear{{Gelman}, {Carlin}, {Stern}, {Dunson}, {Vehtari}  \& {Rubin}}{{Gelman} et~al.}{2014}]{gelman_book}
{Gelman} A.,  {Carlin} J.~B.,  {Stern} H.~S.,  {Dunson} D.~B.,  {Vehtari} A.,   {Rubin} D.~B.,  2014, {Bayesian Data Analysis}, 3rd edn.
Chapman \& Hall/CRC Texts in Statistical Science, CRC Press/Taylor \& Francis

\bibitem[\protect\citeauthoryear{Geyer}{Geyer}{1992}]{geyer92}
Geyer C.~J.,  1992, \mn@doi [Statistical Sci.] {10.1214/ss/1177011137}, 7, 473

\bibitem[\protect\citeauthoryear{Geyer}{Geyer}{2011}]{geyer11}
Geyer C.~J.,  2011, in Brooks S.,  Gelman A.,  Jones G.~L.,   Meng X.-L.,  eds, {Handbook of Markov Chain Monte Carlo}. Chapman \& Hall/CRC, pp 3--48, \url {https://www.mcmchandbook.net/HandbookChapter1.pdf}

\bibitem[\protect\citeauthoryear{{Grayling}, {Thorp}, {Mandel}, {Dhawan}, {Uzsoy}, {Boyd}, {Hayes}  \& {Ward}}{{Grayling} et~al.}{2024}]{Grayling2024}
{Grayling} M.,  {Thorp} S.,  {Mandel} K.~S.,  {Dhawan} S.,  {Uzsoy} A. S.~M.,  {Boyd} B.~M.,  {Hayes} E.~E.,   {Ward} S.~M.,  2024, \mn@doi [\mnras] {10.1093/mnras/stae1202}, \href {https://ui.adsabs.harvard.edu/abs/2024MNRAS.531..953G} {531, 953}

\bibitem[\protect\citeauthoryear{{Gunapati}, {Jain}, {Srijith}  \& {Desai}}{{Gunapati} et~al.}{2022}]{VI_astro_examples}
{Gunapati} G.,  {Jain} A.,  {Srijith} P.~K.,   {Desai} S.,  2022, \mn@doi [\pasa] {10.1017/pasa.2021.64}, \href {https://ui.adsabs.harvard.edu/abs/2022PASA...39....1G} {39, e001}

\bibitem[\protect\citeauthoryear{{Guy}, {Astier}, {Nobili}, {Regnault}  \& {Pain}}{{Guy} et~al.}{2005}]{salt}
{Guy} J.,  {Astier} P.,  {Nobili} S.,  {Regnault} N.,   {Pain} R.,  2005, \mn@doi [\aap] {10.1051/0004-6361:20053025}, \href {https://ui.adsabs.harvard.edu/abs/2005A&A...443..781G} {443, 781}

\bibitem[\protect\citeauthoryear{{Guy} et~al.,}{{Guy} et~al.}{2007}]{salt2}
{Guy} J.,  et~al., 2007, \mn@doi [\aap] {10.1051/0004-6361:20066930}, \href {https://ui.adsabs.harvard.edu/abs/2007A&A...466...11G} {466, 11}

\bibitem[\protect\citeauthoryear{{Hamuy} et~al.,}{{Hamuy} et~al.}{1996a}]{calatololo}
{Hamuy} M.,  et~al., 1996a, \mn@doi [\aj] {10.1086/118192}, \href {https://ui.adsabs.harvard.edu/abs/1996AJ....112.2408H} {112, 2408}

\bibitem[\protect\citeauthoryear{{Hamuy}, {Phillips}, {Suntzeff}, {Schommer}, {Maza}, {Smith}, {Lira}  \& {Aviles}}{{Hamuy} et~al.}{1996b}]{hamuy_templates}
{Hamuy} M.,  {Phillips} M.~M.,  {Suntzeff} N.~B.,  {Schommer} R.~A.,  {Maza} J.,  {Smith} R.~C.,  {Lira} P.,   {Aviles} R.,  1996b, \mn@doi [\aj] {10.1086/118193}, \href {https://ui.adsabs.harvard.edu/abs/1996AJ....112.2438H} {112, 2438}

\bibitem[\protect\citeauthoryear{Han, Liao, Dunson  \& Carin}{Han et~al.}{2016}]{han16}
Han S.,  Liao X.,  Dunson D.,   Carin L.,  2016, in Gretton A.,  Robert C.~C.,  eds,  Proc.\ Machine Learning Res. Vol. 51, Proceedings of the 19th International Conference on Artificial Intelligence and Statistics. PMLR, Cadiz, Spain, pp 829--838 (\mn@eprint {arXiv} {1506.05860})

\bibitem[\protect\citeauthoryear{{Handley}, {Hobson}  \& {Lasenby}}{{Handley} et~al.}{2015}]{polychord}
{Handley} W.~J.,  {Hobson} M.~P.,   {Lasenby} A.~N.,  2015, \mn@doi [\mnras] {10.1093/mnras/stv1911}, \href {https://ui.adsabs.harvard.edu/abs/2015MNRAS.453.4384H} {453, 4384}

\bibitem[\protect\citeauthoryear{{Harris} et~al.,}{{Harris} et~al.}{2020}]{numpy}
{Harris} C.~R.,  et~al., 2020, \mn@doi [\nat] {10.1038/s41586-020-2649-2}, \href {https://ui.adsabs.harvard.edu/abs/2020Natur.585..357H} {585, 357}

\bibitem[\protect\citeauthoryear{{Hatano}, {Branch}  \& {Deaton}}{{Hatano} et~al.}{1998}]{hatano98}
{Hatano} K.,  {Branch} D.,   {Deaton} J.,  1998, \mn@doi [\apj] {10.1086/305903}, \href {https://ui.adsabs.harvard.edu/abs/1998ApJ...502..177H} {502, 177}

\bibitem[\protect\citeauthoryear{{Hicken} et~al.,}{{Hicken} et~al.}{2012}]{hicken12}
{Hicken} M.,  et~al., 2012, \mn@doi [\apjs] {10.1088/0067-0049/200/2/12}, \href {https://ui.adsabs.harvard.edu/abs/2012ApJS..200...12H} {200, 12}

\bibitem[\protect\citeauthoryear{{Hinton} \& {Brout}}{{Hinton} \& {Brout}}{2020}]{pippin}
{Hinton} S.,  {Brout} D.,  2020, \mn@doi [J.\ Open Source Software] {10.21105/joss.02122}, \href {https://ui.adsabs.harvard.edu/abs/2020JOSS....5.2122H} {5, 2122}

\bibitem[\protect\citeauthoryear{Hoffman \& Gelman}{Hoffman \& Gelman}{2014}]{nuts}
Hoffman M.~D.,  Gelman A.,  2014, J.\ Machine Learning Res., \href {https://ui.adsabs.harvard.edu/abs/2011arXiv1111.4246H} {15, 1593}

\bibitem[\protect\citeauthoryear{Hoffman, Blei, Wang  \& Paisley}{Hoffman et~al.}{2013}]{hoffman_svi}
Hoffman M.~D.,  Blei D.~M.,  Wang C.,   Paisley J.,  2013, J.\ Machine Learning Res., \href {https://ui.adsabs.harvard.edu/abs/2012arXiv1206.7051H} {14, 1303}

\bibitem[\protect\citeauthoryear{{Holwerda}, {Reynolds}, {Smith}  \& {Kraan-Korteweg}}{{Holwerda} et~al.}{2015}]{holwerda15}
{Holwerda} B.~W.,  {Reynolds} A.,  {Smith} M.,   {Kraan-Korteweg} R.~C.,  2015, \mn@doi [\mnras] {10.1093/mnras/stu2345}, \href {https://ui.adsabs.harvard.edu/abs/2015MNRAS.446.3768H} {446, 3768}

\bibitem[\protect\citeauthoryear{{Hounsell} et~al.,}{{Hounsell} et~al.}{2018}]{hounsell18}
{Hounsell} R.,  et~al., 2018, \mn@doi [\apj] {10.3847/1538-4357/aac08b}, \href {https://ui.adsabs.harvard.edu/abs/2018ApJ...867...23H} {867, 23}

\bibitem[\protect\citeauthoryear{{Hsiao}}{{Hsiao}}{2009}]{hsiao09}
{Hsiao} E.~Y.,  2009, PhD thesis, Univ. Victoria

\bibitem[\protect\citeauthoryear{{Hsiao}, {Conley}, {Howell}, {Sullivan}, {Pritchet}, {Carlberg}, {Nugent}  \& {Phillips}}{{Hsiao} et~al.}{2007}]{hsiao07}
{Hsiao} E.~Y.,  {Conley} A.,  {Howell} D.~A.,  {Sullivan} M.,  {Pritchet} C.~J.,  {Carlberg} R.~G.,  {Nugent} P.~E.,   {Phillips} M.~M.,  2007, \mn@doi [\apj] {10.1086/518232}, \href {https://ui.adsabs.harvard.edu/abs/2007ApJ...663.1187H} {663, 1187}

\bibitem[\protect\citeauthoryear{Huggins, Kasprzak, Campbell  \& Broderick}{Huggins et~al.}{2020}]{huggins20}
Huggins J.,  Kasprzak M.,  Campbell T.,   Broderick T.,  2020, in Chiappa S.,  Calandra R.,  eds,  Proc.\ Machine Learning Res. Vol. 108, Proceedings of the Twenty Third International Conference on Artificial Intelligence and Statistics. PMLR, pp 1792--1802 (\mn@eprint {arXiv} {1910.04102})

\bibitem[\protect\citeauthoryear{{Hunter}}{{Hunter}}{2007}]{matplotlib}
{Hunter} J.~D.,  2007, \mn@doi [Comput.\ Sci.\ Eng.] {10.1109/MCSE.2007.55}, \href {https://ui.adsabs.harvard.edu/abs/2007CSE.....9...90H} {9, 90}

\bibitem[\protect\citeauthoryear{{Ivezi{\'c}} et~al.,}{{Ivezi{\'c}} et~al.}{2019}]{ivezic19}
{Ivezi{\'c}} {\v{Z}}.,  et~al., 2019, \mn@doi [\apj] {10.3847/1538-4357/ab042c}, \href {https://ui.adsabs.harvard.edu/abs/2019ApJ...873..111I} {873, 111}

\bibitem[\protect\citeauthoryear{{Jha}, {Riess}  \& {Kirshner}}{{Jha} et~al.}{2007}]{MLCS2k2}
{Jha} S.,  {Riess} A.~G.,   {Kirshner} R.~P.,  2007, \mn@doi [\apj] {10.1086/512054}, \href {https://ui.adsabs.harvard.edu/abs/2007ApJ...659..122J} {659, 122}

\bibitem[\protect\citeauthoryear{{Jones} et~al.,}{{Jones} et~al.}{2019}]{jones19}
{Jones} D.~O.,  et~al., 2019, \mn@doi [\apj] {10.3847/1538-4357/ab2bec}, \href {https://ui.adsabs.harvard.edu/abs/2019ApJ...881...19J} {881, 19}

\bibitem[\protect\citeauthoryear{{Jones} et~al.,}{{Jones} et~al.}{2022}]{jones22}
{Jones} D.~O.,  et~al., 2022, \mn@doi [\apj] {10.3847/1538-4357/ac755b}, \href {https://ui.adsabs.harvard.edu/abs/2022ApJ...933..172J} {933, 172}

\bibitem[\protect\citeauthoryear{Jordan, Ghahramani, Jaakkola  \& Saul}{Jordan et~al.}{1999}]{jordan1999introduction}
Jordan M.~I.,  Ghahramani Z.,  Jaakkola T.~S.,   Saul L.~K.,  1999, \mn@doi [Machine Learning] {10.1023/A:1007665907178}, 37, 183

\bibitem[\protect\citeauthoryear{Kantorovich \& Rubinstein}{Kantorovich \& Rubinstein}{1958}]{kr58}
Kantorovich L.,  Rubinstein G.~S.,  1958, Vestnik Leningrad Univ., 13, 52

\bibitem[\protect\citeauthoryear{{Karchev}, {Trotta}  \& {Weniger}}{{Karchev} et~al.}{2023}]{karchev23}
{Karchev} K.,  {Trotta} R.,   {Weniger} C.,  2023, in Machine Learning and the Physical Sciences Workshop, 37th Conference on Neural Information Processing Systems (NeurIPS).  (\mn@eprint {arXiv} {2311.15650})

\bibitem[\protect\citeauthoryear{{Karchev}, {Grayling}, {Boyd}, {Trotta}, {Mandel}  \& {Weniger}}{{Karchev} et~al.}{2024}]{karchev24}
{Karchev} K.,  {Grayling} M.,  {Boyd} B.~M.,  {Trotta} R.,  {Mandel} K.~S.,   {Weniger} C.,  2024, \mn@doi [\mnras] {10.1093/mnras/stae995}, \href {https://ui.adsabs.harvard.edu/abs/2024MNRAS.530.3881K} {530, 3881}

\bibitem[\protect\citeauthoryear{{Kenworthy} et~al.,}{{Kenworthy} et~al.}{2021}]{salt3}
{Kenworthy} W.~D.,  et~al., 2021, \mn@doi [\apj] {10.3847/1538-4357/ac30d8}, \href {https://ui.adsabs.harvard.edu/abs/2021ApJ...923..265K} {923, 265}

\bibitem[\protect\citeauthoryear{{Kessler} \& {Scolnic}}{{Kessler} \& {Scolnic}}{2017}]{bbc}
{Kessler} R.,  {Scolnic} D.,  2017, \mn@doi [\apj] {10.3847/1538-4357/836/1/56}, \href {https://ui.adsabs.harvard.edu/abs/2017ApJ...836...56K} {836, 56}

\bibitem[\protect\citeauthoryear{{Kessler} et~al.,}{{Kessler} et~al.}{2009a}]{snana}
{Kessler} R.,  et~al., 2009a, \mn@doi [\pasp] {10.1086/605984}, \href {https://ui.adsabs.harvard.edu/abs/2009PASP..121.1028K} {121, 1028}

\bibitem[\protect\citeauthoryear{{Kessler} et~al.,}{{Kessler} et~al.}{2009b}]{kessler09}
{Kessler} R.,  et~al., 2009b, \mn@doi [\apjs] {10.1088/0067-0049/185/1/32}, \href {https://ui.adsabs.harvard.edu/abs/2009ApJS..185...32K} {185, 32}

\bibitem[\protect\citeauthoryear{Kingma \& Ba}{Kingma \& Ba}{2015}]{adam_optimizer}
Kingma D.~P.,  Ba J.,  2015, in Bengio Y.,  LeCun Y.,  eds, 3rd International Conference on Learning Representations (ICLR). San Diego, CA, USA (\mn@eprint {arXiv} {1412.6980})

\bibitem[\protect\citeauthoryear{{Kingma} \& {Welling}}{{Kingma} \& {Welling}}{2013}]{svi_vae}
{Kingma} D.~P.,  {Welling} M.,  2013, preprint, \href {https://ui.adsabs.harvard.edu/abs/2013arXiv1312.6114K} {} (\mn@eprint {arXiv} {1312.6114})

\bibitem[\protect\citeauthoryear{{Knox} \& {Millea}}{{Knox} \& {Millea}}{2020}]{hubble_tension_1}
{Knox} L.,  {Millea} M.,  2020, \mn@doi [\prd] {10.1103/PhysRevD.101.043533}, \href {https://ui.adsabs.harvard.edu/abs/2020PhRvD.101d3533K} {101, 043533}

\bibitem[\protect\citeauthoryear{Kolmogorov}{Kolmogorov}{1933}]{kolmogorov33}
Kolmogorov A.,  1933, G.\ Istituto Ital.\ Attuari, 4, 83

\bibitem[\protect\citeauthoryear{{Krisciunas}, {Hastings}, {Loomis}, {McMillan}, {Rest}, {Riess}  \& {Stubbs}}{{Krisciunas} et~al.}{2000}]{krisciunas00}
{Krisciunas} K.,  {Hastings} N.~C.,  {Loomis} K.,  {McMillan} R.,  {Rest} A.,  {Riess} A.~G.,   {Stubbs} C.,  2000, \mn@doi [\apj] {10.1086/309263}, \href {https://ui.adsabs.harvard.edu/abs/2000ApJ...539..658K} {539, 658}

\bibitem[\protect\citeauthoryear{{Krisciunas} et~al.,}{{Krisciunas} et~al.}{2007}]{krisciunas07}
{Krisciunas} K.,  et~al., 2007, \mn@doi [\aj] {10.1086/509126}, \href {https://ui.adsabs.harvard.edu/abs/2007AJ....133...58K} {133, 58}

\bibitem[\protect\citeauthoryear{Kucukelbir, Tran, Ranganath, Gelman  \& Blei}{Kucukelbir et~al.}{2017}]{ADVI}
Kucukelbir A.,  Tran D.,  Ranganath R.,  Gelman A.,   Blei D.~M.,  2017, J.\ Machine Learning Res., \href {https://ui.adsabs.harvard.edu/abs/2016arXiv160300788K} {18, 1}

\bibitem[\protect\citeauthoryear{Kullback \& Leibler}{Kullback \& Leibler}{1951}]{kl_divergence}
Kullback S.,  Leibler R.~A.,  1951, \mn@doi [Ann.\ Math.\ Statistics] {10.1214/aoms/1177729694}, 22, 79

\bibitem[\protect\citeauthoryear{Kumar, Carroll, Hartikainen  \& Martin}{Kumar et~al.}{2019}]{arviz_2019}
Kumar R.,  Carroll C.,  Hartikainen A.,   Martin O.,  2019, \mn@doi [J.\ Open Source Software] {10.21105/joss.01143}, 4, 1143

\bibitem[\protect\citeauthoryear{{LSST Dark Energy Science Collaboration} et~al.,}{{LSST Dark Energy Science Collaboration} et~al.}{2018}]{lsst_info}
{LSST Dark Energy Science Collaboration} et~al., 2018, preprint, \href {https://ui.adsabs.harvard.edu/abs/2018arXiv180901669T} {} (\mn@eprint {arXiv} {1809.01669})

\bibitem[\protect\citeauthoryear{{Lewis} \& {Bridle}}{{Lewis} \& {Bridle}}{2002}]{cosmomc}
{Lewis} A.,  {Bridle} S.,  2002, \mn@doi [\prd] {10.1103/PhysRevD.66.103511}, \href {https://ui.adsabs.harvard.edu/abs/2002PhRvD..66j3511L} {66, 103511}

\bibitem[\protect\citeauthoryear{Liu, McAuliffe, Regier  \& {LSST Dark Energy Science Collaboration}}{Liu et~al.}{2023}]{vi_starfields}
Liu R.,  McAuliffe J.~D.,  Regier J.,   {LSST Dark Energy Science Collaboration} 2023, J.\ Machine Learning Res., \href {https://ui.adsabs.harvard.edu/abs/2021arXiv210202409L} {24, 1}

\bibitem[\protect\citeauthoryear{{Malz}, {Marshall}, {DeRose}, {Graham}, {Schmidt}, {Wechsler}  \& {(LSST Dark Energy Science Collaboration}}{{Malz} et~al.}{2018}]{malz18}
{Malz} A.~I.,  {Marshall} P.~J.,  {DeRose} J.,  {Graham} M.~L.,  {Schmidt} S.~J.,  {Wechsler} R.,   {(LSST Dark Energy Science Collaboration} 2018, \mn@doi [\aj] {10.3847/1538-3881/aac6b5}, \href {https://ui.adsabs.harvard.edu/abs/2018AJ....156...35M} {156, 35}

\bibitem[\protect\citeauthoryear{{Mandel}}{{Mandel}}{2011}]{mandelthesis}
{Mandel} K.~S.,  2011, PhD thesis, Astronomy Department, Harvard Univ.

\bibitem[\protect\citeauthoryear{{Mandel}, {Wood-Vasey}, {Friedman}  \& {Kirshner}}{{Mandel} et~al.}{2009}]{mandel09}
{Mandel} K.~S.,  {Wood-Vasey} W.~M.,  {Friedman} A.~S.,   {Kirshner} R.~P.,  2009, \mn@doi [\apj] {10.1088/0004-637X/704/1/629}, \href {https://ui.adsabs.harvard.edu/abs/2009ApJ...704..629M} {704, 629}

\bibitem[\protect\citeauthoryear{{Mandel}, {Narayan}  \& {Kirshner}}{{Mandel} et~al.}{2011}]{mandel11}
{Mandel} K.~S.,  {Narayan} G.,   {Kirshner} R.~P.,  2011, \mn@doi [\apj] {10.1088/0004-637X/731/2/120}, \href {https://ui.adsabs.harvard.edu/abs/2011ApJ...731..120M} {731, 120}

\bibitem[\protect\citeauthoryear{{Mandel}, {Scolnic}, {Shariff}, {Foley}  \& {Kirshner}}{{Mandel} et~al.}{2017}]{mandel17}
{Mandel} K.~S.,  {Scolnic} D.~M.,  {Shariff} H.,  {Foley} R.~J.,   {Kirshner} R.~P.,  2017, \mn@doi [\apj] {10.3847/1538-4357/aa6038}, \href {https://ui.adsabs.harvard.edu/abs/2017ApJ...842...93M} {842, 93}

\bibitem[\protect\citeauthoryear{{Mandel}, {Thorp}, {Narayan}, {Friedman}  \& {Avelino}}{{Mandel} et~al.}{2022}]{mandel22}
{Mandel} K.~S.,  {Thorp} S.,  {Narayan} G.,  {Friedman} A.~S.,   {Avelino} A.,  2022, \mn@doi [\mnras] {10.1093/mnras/stab3496}, \href {https://ui.adsabs.harvard.edu/abs/2022MNRAS.510.3939M} {510, 3939}

\bibitem[\protect\citeauthoryear{{Margossian}, {Pillaud-Vivien}  \& {Saul}}{{Margossian} et~al.}{2024}]{margossian2024}
{Margossian} C.~C.,  {Pillaud-Vivien} L.,   {Saul} L.~K.,  2024, preprint, \href {https://ui.adsabs.harvard.edu/abs/2024arXiv240313748M} {} (\mn@eprint {arXiv} {2403.13748})

\bibitem[\protect\citeauthoryear{{Massey Jr.}}{{Massey Jr.}}{1951}]{massey1951kolmogorov}
{Massey Jr.} F.~J.,  1951, J.\ American Statistical Association, 46, 68

\bibitem[\protect\citeauthoryear{Minka}{Minka}{2001}]{minka2013expectation}
Minka T.~P.,  2001, in Breese J.~S.,  Koller D.,  eds, {UAI} '01: Proceedings of the 17th Conference in Uncertainty in Artificial Intelligence, University of Washington, Seattle, Washington, USA, August 2-5, 2001. Morgan Kaufmann, pp 362--369 (\mn@eprint {arXiv} {1301.2294})

\bibitem[\protect\citeauthoryear{Minka}{Minka}{2005}]{minka2005divergence}
Minka T.,  2005, Technical Report MSR-TR-2005-173, Divergence measures and message passing.
Microsoft Research, \url{https://tminka.github.io/papers/message-passing/}

\bibitem[\protect\citeauthoryear{Papamakarios, Pavlakou  \& Murray}{Papamakarios et~al.}{2017}]{papamakarios17}
Papamakarios G.,  Pavlakou T.,   Murray I.,  2017, in Guyon I.,  Luxburg U.~V.,  Bengio S.,  Wallach H.,  Fergus R.,  Vishwanathan S.,   Garnett R.,  eds,  Advances in Neural Information Processing Systems Vol. 30, NIPS 2017. Curran Associates, Inc. (\mn@eprint {arXiv} {1705.07057})

\bibitem[\protect\citeauthoryear{{Patil}, {Bovy}, {Jaimungal}, {Frankel}  \& {Leung}}{{Patil} et~al.}{2023}]{patil23}
{Patil} A.~A.,  {Bovy} J.,  {Jaimungal} S.,  {Frankel} N.,   {Leung} H.~W.,  2023, \mn@doi [\mnras] {10.1093/mnras/stad2820}, \href {https://ui.adsabs.harvard.edu/abs/2023MNRAS.526.1997P} {526, 1997}

\bibitem[\protect\citeauthoryear{Perepolkin, Goodrich  \& Sahlin}{Perepolkin et~al.}{2023}]{perepolkin23}
Perepolkin D.,  Goodrich B.,   Sahlin U.,  2023, \mn@doi [Comput.\ Statistics Data Analysis] {10.1016/j.csda.2023.107795}, 187, 107795

\bibitem[\protect\citeauthoryear{{Perlmutter} et~al.,}{{Perlmutter} et~al.}{1999}]{Perlmutter_1999}
{Perlmutter} S.,  et~al., 1999, \mn@doi [\apj] {10.1086/307221}, \href {https://ui.adsabs.harvard.edu/abs/1999ApJ...517..565P} {517, 565}

\bibitem[\protect\citeauthoryear{{Peterson} et~al.,}{{Peterson} et~al.}{2023}]{peterson23}
{Peterson} E.~R.,  et~al., 2023, \mn@doi [\mnras] {10.1093/mnras/stad1077}, \href {https://ui.adsabs.harvard.edu/abs/2023MNRAS.522.2478P} {522, 2478}

\bibitem[\protect\citeauthoryear{{Phan}, {Pradhan}  \& {Jankowiak}}{{Phan} et~al.}{2019}]{numpyro}
{Phan} D.,  {Pradhan} N.,   {Jankowiak} M.,  2019, preprint, \href {https://ui.adsabs.harvard.edu/abs/2019arXiv191211554P} {} (\mn@eprint {arXiv} {1912.11554})

\bibitem[\protect\citeauthoryear{{Phillips}}{{Phillips}}{1993}]{phillips}
{Phillips} M.~M.,  1993, \mn@doi [\apjl] {10.1086/186970}, \href {https://ui.adsabs.harvard.edu/abs/1993ApJ...413L.105P} {413, L105}

\bibitem[\protect\citeauthoryear{{Phillips}, {Lira}, {Suntzeff}, {Schommer}, {Hamuy}  \& {Maza}}{{Phillips} et~al.}{1999}]{phillips99}
{Phillips} M.~M.,  {Lira} P.,  {Suntzeff} N.~B.,  {Schommer} R.~A.,  {Hamuy} M.,   {Maza} J.,  1999, \mn@doi [\aj] {10.1086/301032}, \href {https://ui.adsabs.harvard.edu/abs/1999AJ....118.1766P} {118, 1766}

\bibitem[\protect\citeauthoryear{{Phillips} et~al.,}{{Phillips} et~al.}{2019}]{csp2}
{Phillips} M.~M.,  et~al., 2019, \mn@doi [\pasp] {10.1088/1538-3873/aae8bd}, \href {https://ui.adsabs.harvard.edu/abs/2019PASP..131a4001P} {131, 014001}

\bibitem[\protect\citeauthoryear{{Pierel} et~al.,}{{Pierel} et~al.}{2024}]{pierel24}
{Pierel} J.~D.~R.,  et~al., 2024, \mn@doi [\apj] {10.3847/1538-4357/ad3c43}, \href {https://ui.adsabs.harvard.edu/abs/2024ApJ...967...50P} {967, 50}

\bibitem[\protect\citeauthoryear{{Planck Collaboration} et~al.,}{{Planck Collaboration} et~al.}{2020}]{planck}
{Planck Collaboration} et~al., 2020, \mn@doi [\aap] {10.1051/0004-6361/201833910}, \href {https://ui.adsabs.harvard.edu/abs/2020A&A...641A...6P} {641, A6}

\bibitem[\protect\citeauthoryear{{Popovic}, {Brout}, {Kessler}, {Scolnic}  \& {Lu}}{{Popovic} et~al.}{2021}]{popovic21}
{Popovic} B.,  {Brout} D.,  {Kessler} R.,  {Scolnic} D.,   {Lu} L.,  2021, \mn@doi [\apj] {10.3847/1538-4357/abf14f}, \href {https://ui.adsabs.harvard.edu/abs/2021ApJ...913...49P} {913, 49}

\bibitem[\protect\citeauthoryear{{Pskovskii}}{{Pskovskii}}{1977}]{pskovskii77}
{Pskovskii} I.~P.,  1977, \sovast, \href {https://ui.adsabs.harvard.edu/abs/1977SvA....21..675P} {21, 675}

\bibitem[\protect\citeauthoryear{Ranganath, Gerrish  \& Blei}{Ranganath et~al.}{2014}]{bbvi}
Ranganath R.,  Gerrish S.,   Blei D.,  2014, in Kaski S.,  Corander J.,  eds,  Proc.\ Machine Learning Res. Vol. 33, Proceedings of the Seventeenth International Conference on Artificial Intelligence and Statistics. PMLR, Reykjavik, Iceland, pp 814--822 (\mn@eprint {arXiv} {1401.0118})

\bibitem[\protect\citeauthoryear{Rasmussen \& Williams}{Rasmussen \& Williams}{2006}]{Rasmussen_Williams}
Rasmussen C.~E.,  Williams C. K.~I.,  2006, Gaussian Processes for Machine Learning.
Adaptive Computation \& Machine Learning, MIT Press

\bibitem[\protect\citeauthoryear{Regier, Miller, Schlegel, Adams, McAuliffe  \& Prabhat}{Regier et~al.}{2019}]{prabhat}
Regier J.,  Miller A.~C.,  Schlegel D.,  Adams R.~P.,  McAuliffe J.~D.,   Prabhat 2019, \mn@doi [Ann.\ Applied Statistics] {10.1214/19-AOAS1258}, \href {https://ui.adsabs.harvard.edu/abs/2018arXiv180300113R} {13, 1884 }

\bibitem[\protect\citeauthoryear{{Rest} et~al.,}{{Rest} et~al.}{2014}]{rest14}
{Rest} A.,  et~al., 2014, \mn@doi [\apj] {10.1088/0004-637X/795/1/44}, \href {https://ui.adsabs.harvard.edu/abs/2014ApJ...795...44R} {795, 44}

\bibitem[\protect\citeauthoryear{Rezende \& Mohamed}{Rezende \& Mohamed}{2015}]{rezende15}
Rezende D.,  Mohamed S.,  2015, in Bach F.,  Blei D.,  eds,  Proc.\ Machine Learning Res. Vol. 37, Proceedings of the 32nd International Conference on Machine Learning. PMLR, Lille, France, pp 1530--1538 (\mn@eprint {arXiv} {1505.05770})

\bibitem[\protect\citeauthoryear{{Riello} \& {Patat}}{{Riello} \& {Patat}}{2005}]{riello05}
{Riello} M.,  {Patat} F.,  2005, \mn@doi [\mnras] {10.1111/j.1365-2966.2005.09348.x}, \href {https://ui.adsabs.harvard.edu/abs/2005MNRAS.362..671R} {362, 671}

\bibitem[\protect\citeauthoryear{{Riess}, {Press}  \& {Kirshner}}{{Riess} et~al.}{1995}]{LCS}
{Riess} A.~G.,  {Press} W.~H.,   {Kirshner} R.~P.,  1995, \mn@doi [\apjl] {10.1086/187704}, \href {https://ui.adsabs.harvard.edu/abs/1995ApJ...438L..17R} {438, L17}

\bibitem[\protect\citeauthoryear{{Riess}, {Press}  \& {Kirshner}}{{Riess} et~al.}{1996a}]{MLCS}
{Riess} A.~G.,  {Press} W.~H.,   {Kirshner} R.~P.,  1996a, \mn@doi [\apj] {10.1086/178129}, \href {https://ui.adsabs.harvard.edu/abs/1996ApJ...473...88R} {473, 88}

\bibitem[\protect\citeauthoryear{{Riess}, {Press}  \& {Kirshner}}{{Riess} et~al.}{1996b}]{riess96_dust}
{Riess} A.~G.,  {Press} W.~H.,   {Kirshner} R.~P.,  1996b, \mn@doi [\apj] {10.1086/178174}, \href {https://ui.adsabs.harvard.edu/abs/1996ApJ...473..588R} {473, 588}

\bibitem[\protect\citeauthoryear{{Riess} et~al.,}{{Riess} et~al.}{1998}]{Riess_1998}
{Riess} A.~G.,  et~al., 1998, \mn@doi [\aj] {10.1086/300499}, \href {https://ui.adsabs.harvard.edu/abs/1998AJ....116.1009R} {116, 1009}

\bibitem[\protect\citeauthoryear{{Riess} et~al.,}{{Riess} et~al.}{1999}]{riess99}
{Riess} A.~G.,  et~al., 1999, \mn@doi [\aj] {10.1086/300738}, \href {https://ui.adsabs.harvard.edu/abs/1999AJ....117..707R} {117, 707}

\bibitem[\protect\citeauthoryear{{Riess} et~al.,}{{Riess} et~al.}{2016}]{riess2016}
{Riess} A.~G.,  et~al., 2016, \mn@doi [\apj] {10.3847/0004-637X/826/1/56}, \href {https://ui.adsabs.harvard.edu/abs/2016ApJ...826...56R} {826, 56}

\bibitem[\protect\citeauthoryear{{Riess} et~al.,}{{Riess} et~al.}{2022}]{riess22}
{Riess} A.~G.,  et~al., 2022, \mn@doi [\apjl] {10.3847/2041-8213/ac5c5b}, \href {https://ui.adsabs.harvard.edu/abs/2022ApJ...934L...7R} {934, L7}

\bibitem[\protect\citeauthoryear{{Rizzato} \& {Sellentin}}{{Rizzato} \& {Sellentin}}{2023}]{vi_precision_cosmology}
{Rizzato} M.,  {Sellentin} E.,  2023, \mn@doi [\mnras] {10.1093/mnras/stad638}, \href {https://ui.adsabs.harvard.edu/abs/2023MNRAS.521.1152R} {521, 1152}

\bibitem[\protect\citeauthoryear{{Rose} et~al.,}{{Rose} et~al.}{2021}]{rose_roman}
{Rose} B.~M.,  et~al., 2021, preprint, \href {https://ui.adsabs.harvard.edu/abs/2021arXiv211103081R} {} (\mn@eprint {arXiv} {2111.03081})

\bibitem[\protect\citeauthoryear{{Rust}}{{Rust}}{1975}]{rust75}
{Rust} B.~W.,  1975, Bull.\ American Astron.\ Soc., \href {https://ui.adsabs.harvard.edu/abs/1975BAAS....7..236R} {7, 236}

\bibitem[\protect\citeauthoryear{{Salim} \& {Narayanan}}{{Salim} \& {Narayanan}}{2020}]{salim20}
{Salim} S.,  {Narayanan} D.,  2020, \mn@doi [\araa] {10.1146/annurev-astro-032620-021933}, \href {https://ui.adsabs.harvard.edu/abs/2020ARA&A..58..529S} {58, 529}

\bibitem[\protect\citeauthoryear{S{\'a}nchez, Cabrera, Huijse  \& F{\"o}rster}{S{\'a}nchez et~al.}{2021}]{sanchez2021amortized}
S{\'a}nchez A.,  Cabrera G.,  Huijse P.,   F{\"o}rster F.,  2021, in Machine Learning and the Physical Sciences Workshop, 35th Conference on Neural Information Processing Systems (NeurIPS). \url {https://ml4physicalsciences.github.io/2021/files/NeurIPS_ML4PS_2021_10.pdf}

\bibitem[\protect\citeauthoryear{{Sandage} \& {Tammann}}{{Sandage} \& {Tammann}}{1993}]{sandage93}
{Sandage} A.,  {Tammann} G.~A.,  1993, \mn@doi [\apj] {10.1086/173137}, \href {https://ui.adsabs.harvard.edu/abs/1993ApJ...415....1S} {415, 1}

\bibitem[\protect\citeauthoryear{{Saunders} et~al.,}{{Saunders} et~al.}{2018}]{SNEMO}
{Saunders} C.,  et~al., 2018, \mn@doi [\apj] {10.3847/1538-4357/aaec7e}, \href {https://ui.adsabs.harvard.edu/abs/2018ApJ...869..167S} {869, 167}

\bibitem[\protect\citeauthoryear{{Schlafly} \& {Finkbeiner}}{{Schlafly} \& {Finkbeiner}}{2011}]{galaxy_dust}
{Schlafly} E.~F.,  {Finkbeiner} D.~P.,  2011, \mn@doi [\apj] {10.1088/0004-637X/737/2/103}, \href {https://ui.adsabs.harvard.edu/abs/2011ApJ...737..103S} {737, 103}

\bibitem[\protect\citeauthoryear{{Schlafly} et~al.,}{{Schlafly} et~al.}{2016}]{schlafly2}
{Schlafly} E.~F.,  et~al., 2016, \mn@doi [\apj] {10.3847/0004-637X/821/2/78}, \href {https://ui.adsabs.harvard.edu/abs/2016ApJ...821...78S} {821, 78}

\bibitem[\protect\citeauthoryear{{Schlegel}, {Finkbeiner}  \& {Davis}}{{Schlegel} et~al.}{1998}]{sfd98}
{Schlegel} D.~J.,  {Finkbeiner} D.~P.,   {Davis} M.,  1998, \mn@doi [\apj] {10.1086/305772}, \href {https://ui.adsabs.harvard.edu/abs/1998ApJ...500..525S} {500, 525}

\bibitem[\protect\citeauthoryear{{Scolnic} et~al.,}{{Scolnic} et~al.}{2018}]{scolnic18}
{Scolnic} D.~M.,  et~al., 2018, \mn@doi [\apj] {10.3847/1538-4357/aab9bb}, \href {https://ui.adsabs.harvard.edu/abs/2018ApJ...859..101S} {859, 101}

\bibitem[\protect\citeauthoryear{Smirnov}{Smirnov}{1948}]{smirnov48}
Smirnov N.,  1948, \mn@doi [Ann.\ Math.\ Statistics] {10.1214/aoms/1177730256}, 19, 279

\bibitem[\protect\citeauthoryear{{Speagle}}{{Speagle}}{2020}]{dynesty}
{Speagle} J.~S.,  2020, \mn@doi [\mnras] {10.1093/mnras/staa278}, \href {https://ui.adsabs.harvard.edu/abs/2020MNRAS.493.3132S} {493, 3132}

\bibitem[\protect\citeauthoryear{{Stan Development Team}}{{Stan Development Team}}{2024}]{stan}
{Stan Development Team} 2024, Stan Modelling Language Users Guide and Reference Manual v.2.34.
\url {https://mc-stan.org}

\bibitem[\protect\citeauthoryear{{Talts}, {Betancourt}, {Simpson}, {Vehtari}  \& {Gelman}}{{Talts} et~al.}{2018}]{talts18}
{Talts} S.,  {Betancourt} M.,  {Simpson} D.,  {Vehtari} A.,   {Gelman} A.,  2018, preprint, \href {https://ui.adsabs.harvard.edu/abs/2018arXiv180406788T} {} (\mn@eprint {arXiv} {1804.06788})

\bibitem[\protect\citeauthoryear{{Thorp} \& {Mandel}}{{Thorp} \& {Mandel}}{2022}]{thorp22}
{Thorp} S.,  {Mandel} K.~S.,  2022, \mn@doi [\mnras] {10.1093/mnras/stac2714}, \href {https://ui.adsabs.harvard.edu/abs/2022MNRAS.517.2360T} {517, 2360}

\bibitem[\protect\citeauthoryear{{Thorp}, {Mandel}, {Jones}, {Ward}  \& {Narayan}}{{Thorp} et~al.}{2021}]{thorp21}
{Thorp} S.,  {Mandel} K.~S.,  {Jones} D.~O.,  {Ward} S.~M.,   {Narayan} G.,  2021, \mn@doi [\mnras] {10.1093/mnras/stab2849}, \href {https://ui.adsabs.harvard.edu/abs/2021MNRAS.508.4310T} {508, 4310}

\bibitem[\protect\citeauthoryear{{Thorp}, {Mandel}, {Jones}, {Kirshner}  \& {Challis}}{{Thorp} et~al.}{2024}]{thorp2024}
{Thorp} S.,  {Mandel} K.~S.,  {Jones} D.~O.,  {Kirshner} R.~P.,   {Challis} P.~M.,  2024, \mn@doi [\mnras] {10.1093/mnras/stae1111}, \href {https://ui.adsabs.harvard.edu/abs/2024MNRAS.530.4016T} {530, 4016}

\bibitem[\protect\citeauthoryear{Tran, Blei  \& Airoldi}{Tran et~al.}{2015}]{tran15}
Tran D.,  Blei D.,   Airoldi E.~M.,  2015, in Cortes C.,  Lawrence N.,  Lee D.,  Sugiyama M.,   Garnett R.,  eds,  Advances in Neural Information Processing Systems Vol. 28, NIPS 2015. Curran Associates, Inc. (\mn@eprint {arXiv} {1506.03159})

\bibitem[\protect\citeauthoryear{{Tripp}}{{Tripp}}{1998}]{tripp98}
{Tripp} R.,  1998, \aap, \href {https://ui.adsabs.harvard.edu/abs/1998A&A...331..815T} {331, 815}

\bibitem[\protect\citeauthoryear{Turner \& Sahani}{Turner \& Sahani}{2011}]{turner2011two}
Turner R.~E.,  Sahani M.,  2011, in Barber D.,  Cemgil A.~T.,   Chiappa S.,  eds, Bayesian Time Series Models. Cambridge University Press, pp 104--124, \url {https://doi.org/10.1017/CBO9780511984679.006}

\bibitem[\protect\citeauthoryear{{Uzsoy}}{{Uzsoy}}{2022}]{Uzsoy22}
{Uzsoy} A.~S.,  2022, Master's thesis, Univ.\ Cambridge, \url {https://www.mlmi.eng.cam.ac.uk/files/2021-2022_dissertations/scalable_bayesian_inference_for_probabilistic_spectrotemporal_models.pdf}

\bibitem[\protect\citeauthoryear{Vaserstein}{Vaserstein}{1969}]{wasserstein}
Vaserstein L.~N.,  1969, Problemy Peredachi Inf., 5, 64

\bibitem[\protect\citeauthoryear{Vehtari, Gelman, Simpson, Carpenter  \& B{\"u}rkner}{Vehtari et~al.}{2021}]{vehtari21}
Vehtari A.,  Gelman A.,  Simpson D.,  Carpenter B.,   B{\"u}rkner P.-C.,  2021, \mn@doi [Bayesian Analysis] {10.1214/20-BA1221}, 16, 667

\bibitem[\protect\citeauthoryear{{Vehtari}, {Simpson}, {Gelman}, {Yao}  \& {Gabry}}{{Vehtari} et~al.}{2024}]{vehtari15}
{Vehtari} A.,  {Simpson} D.,  {Gelman} A.,  {Yao} Y.,   {Gabry} J.,  2024, \mn@doi [J.\ Machine Learning Res.] {10.48550/arXiv.1507.02646}, \href {https://ui.adsabs.harvard.edu/abs/2015arXiv150702646V} {25, 1}

\bibitem[\protect\citeauthoryear{Villar}{Villar}{2022}]{villar2022}
Villar V.~A.,  2022, in Machine Learning and the Physical Sciences Workshop, 36th Conference on Neural Information Processing Systems (NeurIPS).  (\mn@eprint {arXiv} {2211.04480})

\bibitem[\protect\citeauthoryear{{Villar} et~al.,}{{Villar} et~al.}{2020}]{villar20}
{Villar} V.~A.,  et~al., 2020, \mn@doi [\apj] {10.3847/1538-4357/abc6fd}, \href {https://ui.adsabs.harvard.edu/abs/2020ApJ...905...94V} {905, 94}

\bibitem[\protect\citeauthoryear{{Villar}, {Cranmer}, {Berger}, {Contardo}, {Ho}, {Hosseinzadeh}  \& {Lin}}{{Villar} et~al.}{2021}]{villar_vae_classifier}
{Villar} V.~A.,  {Cranmer} M.,  {Berger} E.,  {Contardo} G.,  {Ho} S.,  {Hosseinzadeh} G.,   {Lin} J. Y.-Y.,  2021, \mn@doi [\apjs] {10.3847/1538-4365/ac0893}, \href {https://ui.adsabs.harvard.edu/abs/2021ApJS..255...24V} {255, 24}

\bibitem[\protect\citeauthoryear{Virtanen et~al.,}{Virtanen et~al.}{2020}]{2020SciPy-NMeth}
Virtanen P.,  et~al., 2020, \mn@doi [Nature Methods] {10.1038/s41592-019-0686-2}, \href {https://rdcu.be/b08Wh} {17, 261}

\bibitem[\protect\citeauthoryear{Wainwright \& Jordan}{Wainwright \& Jordan}{2008}]{wainwright2008graphical}
Wainwright M.~J.,  Jordan M.~I.,  2008, \mn@doi [Foundations Trends Machine Learning] {10.1561/2200000001}, 1, 1

\bibitem[\protect\citeauthoryear{{Ward}, {Dhawan}, {Mandel}, {Grayling}  \& {Thorp}}{{Ward} et~al.}{2023a}]{birdsnack}
{Ward} S.~M.,  {Dhawan} S.,  {Mandel} K.~S.,  {Grayling} M.,   {Thorp} S.,  2023a, \mn@doi [\mnras] {10.1093/mnras/stad3159}, \href {https://ui.adsabs.harvard.edu/abs/2023MNRAS.526.5715W} {526, 5715}

\bibitem[\protect\citeauthoryear{{Ward} et~al.,}{{Ward} et~al.}{2023b}]{ward22}
{Ward} S.~M.,  et~al., 2023b, \mn@doi [\apj] {10.3847/1538-4357/acf7bb}, \href {https://ui.adsabs.harvard.edu/abs/2023ApJ...956..111W} {956, 111}

\bibitem[\protect\citeauthoryear{{Wingate} \& {Weber}}{{Wingate} \& {Weber}}{2013}]{svi_ppl}
{Wingate} D.,  {Weber} T.,  2013, preprint, \href {https://ui.adsabs.harvard.edu/abs/2013arXiv1301.1299W} {} (\mn@eprint {arXiv} {1301.1299})

\bibitem[\protect\citeauthoryear{{Wojtak}, {Hjorth}  \& {Hjortlund}}{{Wojtak} et~al.}{2023}]{wojtak23}
{Wojtak} R.,  {Hjorth} J.,   {Hjortlund} J.~O.,  2023, \mn@doi [\mnras] {10.1093/mnras/stad2590}, \href {https://ui.adsabs.harvard.edu/abs/2023MNRAS.525.5187W} {525, 5187}

\bibitem[\protect\citeauthoryear{Yao, Vehtari, Simpson  \& Gelman}{Yao et~al.}{2018}]{yesbutdiditwork}
Yao Y.,  Vehtari A.,  Simpson D.,   Gelman A.,  2018, in Dy J.,  Krause A.,  eds,  Proc.\ Machine Learning Res. Vol. 80, Proceedings of the 35th International Conference on Machine Learning. PMLR, pp 5581--5590 (\mn@eprint {arXiv} {1802.02538})

\makeatother
\end{thebibliography}



\appendix
\section{Marginal Densities of the MVZLTN}
\label{zltn_marginals}

In this appendix, we present analytic results for the marginal probability densities of subsets of $M<N$ variables of the $N$-dimensional MVZLTN distribution defined in Section \ref{zltn_section}. For the results in this section, we will consider an $N$-vector $\bm{\phi}=(\phi_t, \bm{\phi}_u)^\top$, with truncated variable $\phi_t$, and $N-1$ untruncated variables $\bm{\phi}_u$. The full $\bm{\phi}$ vector will have an MVZLTN density,
\begin{align}
    P(\bm{\phi}) &= \mathrm{MVZLTN}(\bm{\phi}|\bm{\mu}, \bm{\Sigma}) \nonumber\\ &= \mathrm{ZLTN}( \phi_t |\, \mu_t, \sigma_t^2) \times \mathrm{N}( \bm{\phi}_u |\,\bm{\mu}_{u|t}, \bm{\Sigma}_{uu|t} ),
\end{align}
with the mean and covariance parameters partitioned as in Equation \ref{eq:zltn_params}, and with the conditional mean and covariance of $\bm{\phi}_u|\phi_t$ given by Equations \ref{eq:conditional_mean} and \ref{eq:conditional_cov}.

In the following subsections, we will derive all univariate and bivariate marginals. We have validated numerical implementations of these analytic expressions against Monte Carlo samples from the MVZLTN joint distribution. Marginal densities over $2<M<N$ variables are straightforward generalizations of these results.

\subsection{Univariate Marginal of the Truncated Variable}
By definition, the marginal density of the truncated variable $\phi_t$ is just a univariate ZLTN distribution,
\begin{equation}
    P(\phi_t) = \mathrm{ZLTN}(\phi_t|\mu_t, \sigma_t^2),
\end{equation}
as defined in Equation \ref{eq:sample_zltn}.

\subsection{Bivariate Marginal of the Truncated Variable and an Untruncated Variable}
Marginalization over any number of untruncated variables can be achieved trivially, and yields a bivariate ZLTN. Marginalizing over all but the $i$th untruncated variables gives
\begin{equation}
    P(\phi_t, \phi_u^i) = \mathrm{ZLTN}(\phi_t |\, \mu_t, \sigma_t^2) \times \mathrm{N}(\phi_u^i | \, \mu_{u|t}^i, \Sigma_{uu|t}^{i,i} ),
    \label{eq:biv_tu}
\end{equation}
where in the second term we have picked out the $i$th element of the conditional mean (Eq.\ \ref{eq:conditional_mean}) and the $(i,i)$th element of the conditional covariance matrix (Eq.\ \ref{eq:conditional_cov}). 

\subsection{Univariate Marginal of an Untruncated Variable}
From Eq.\ \ref{eq:biv_tu}, we can derive the marginal of the $i$th untruncated variable, $\phi_u^i$, by integrating over the truncated variable, $\phi_t$. We begin by expanding the first term in Eq.\ \ref{eq:biv_tu} using the definition of the ZLTN (Eq.\ \ref{eq:sample_zltn}), expanding $\mu_{u|t}^i$ using Eq.\ \ref{eq:conditional_mean}, and rewriting the second term as a Gaussian density over $\phi_t$:
\begin{equation}
\begin{split}
P(\phi_u^i) &= \int_0^\infty \mathrm{ZLTN}(\phi_t |\, \mu_t, \sigma_t^2) \times \mathrm{N}(\phi_u^i | \, \mu_{u|t}^i, \Sigma_{uu|t}^{i,i} ) \, d\phi_t \\
&= \frac{1}{\Psi(\mu_t/\sigma_t)} \int_0^\infty \mathrm{N}(\phi_t |\, \mu_t, \sigma_t^2) \times \mathrm{N}(\phi_u^i | \, \mu_{u|t}^i, \Sigma_{uu|t}^{i,i} ) \,d\phi_t \\
\end{split}
\end{equation}
First, we handle the case of $\Sigma_{ut}^i \ne 0$. The preceding expression can be written as:
\begin{equation}
P(\phi_u^i)= \frac{1}{\Psi(\mu_t/\sigma_t)} \int_0^\infty \mathrm{N}(\phi_t |\, \mu_t, \sigma_t^2) \times \frac{\sigma_t^2}{\Sigma_{ut}^i} \mathrm{N}(y|\, \phi_t, s_t^2) \, d\phi_t, \\
\label{eq:zltnexpandstep}
\end{equation}
where $y = \mu_t +  \sigma_t^2\times(\phi_u^i - \mu_u^i)/\Sigma_{ut}^i$
and $s_t^2 = \Sigma_{uu|t}^{i,i} \times \sigma_t^4 / (\Sigma_{ut}^i)^2$. We take the product of the two Gaussian densities, giving
\begin{equation}
P(\phi_u^i) = \frac{\sigma_t^2\times \mathrm{N}(y |\, \mu_t, \sigma_t^2 + s_t^2)}{\Sigma_{ut}^i\times\Psi(\mu_t/\sigma_t)} \int_0^\infty \mathrm{N}(\phi_t |\, \tilde{\mu}_t, \tilde{\sigma}_t^2) \, d\phi_t,
\label{eq:gaussproduct}
\end{equation}
where $\tilde{\mu}_t = \tilde{\sigma}_t^2 ( \sigma_t^{-2} \mu_t + s_t^{-2} y )$ and $\tilde{\sigma}_t^2 = [ \sigma_t^{-2} + s_t^{-2}]^{-1}$. Notice that the variable of interest $\phi_u^i$ is contained in $y$ and therefore $\tilde{\mu}_t$. The remaining integral will then have the solution $[1-\Psi(-\tilde{\mu}_t / \tilde{\sigma}_t)] = \Psi(\tilde{\mu}_t / \tilde{\sigma}_t)$, giving
\begin{equation}
P(\phi_u^i) = \frac{\sigma_t^2 \times \Psi(\tilde{\mu}_t / \tilde{\sigma}_t)}{\Sigma_{ut}^i \times \Psi(\mu_t/\sigma_t)} \times \mathrm{N}(y |\, \mu_t, \sigma_t^2 + s_t^2).
\label{eq:uni_u}
\end{equation}

In the case of $\Sigma_{ut}^i = 0$, the marginalization simplifies considerably to:
\begin{equation}
P(\phi_u^i) = N(\phi_u^i |\, \mu_u^i, \Sigma_{uu}^{i,i}).
\end{equation}

\subsection{Bivariate Marginal of Two Untruncated Variables}
Finally, we will derive the bivariate joint density over the $i$th and $j$th untruncated variables, which we will include in a reduced untruncated vector $\bm{\phi}_u' = (\phi_u^i, \phi_u^j)^\top$. Similarly, we will let $\bm{\mu}_u' = (\mu_u^i, \mu_u^j)^\top$ and $\bm{\Sigma}_{ut}' = (\bm{\Sigma}_{ut}^i, \bm{\Sigma}_{ut}^j )^\top$, and will define a $2 \times 2$ submatrix of the full conditional covariance,
\begin{equation}
\bm{\Sigma}_{uu|t}' = \begin{pmatrix} \Sigma_{uu|t}^{i,i} & \Sigma_{uu|t}^{j,i} \\ \Sigma_{uu|t}^{i,j} & \Sigma_{uu|t}^{j,j} \end{pmatrix}.
\end{equation}
Analogously to Equation \ref{eq:biv_tu}, the trivariate marginal density over $\phi_t$ and $\bm{\phi}_u'$ will be given by
\begin{equation}
        P(\phi_t, \bm{\phi}_u') = \mathrm{ZLTN}( \phi_t | \, \mu_t, \sigma_t^2) \times \mathrm{N}\left(\bm{\phi}_u' \Bigg|\, \bm{\mu}_u'  + \bm{\Sigma}_{ut}'   \frac{\phi_t - \mu_t}{\sigma_t^2}, \bm{\Sigma}_{uu|t}'   \right).
\end{equation}
First, we handle the case of $\bm{\Sigma}_{ut}' \ne \bm{0}$. The preceding expression can be written as:
\begin{equation}
        P(\phi_t, \bm{\phi}_u') =  \mathrm{ZLTN}( \phi_t | \, \mu_t, \sigma_t^2) \times \mathrm{N}(\bm{y} |\, \bm{x} \phi_t, \bm{W}),
    \label{eq:trivariate}
\end{equation}
where we have explicitly expanded out the conditional mean (Eq.\ \ref{eq:conditional_mean}), and define for convenience: $\bm{y} = \bm{\phi}_u' -  \bm{\mu}_u' + \bm{\Sigma}_{ut}' \mu_t/\sigma_t^2$, $\bm{x} = \bm{\Sigma}_{ut}' /\sigma_t^2$, and $\bm{W} = \bm{\Sigma}_{uu|t}'$. We can recognise the second term in Equation \ref{eq:trivariate} as being the likelihood for $\phi_t$ in a linear model, $\bm{y}=\phi_t \bm{x}$, with correlated Gaussian errors on $\bm{y}$ with covariance $\bm{W}$ (see e.g.\ \citealp{gelman_book}, eq.\ 14.11). We can separate out the $\phi_t$-dependent part of this term to obtain
\begin{equation}
    \mathrm{N}(\bm{y} |\, \bm{x} \phi_t, \bm{W}) = \sqrt{2\pi s_t^2} \times \mathrm{N}(\phi_t |\, \hat{\phi}_t, s^2_t) \times \mathrm{N}(\bm{y} | \, \bm{x} \hat{\phi}_t, \bm{W}),
    \label{eq:linearmle}
\end{equation}
where $\hat{\phi}_t = s_t^2 \, \bm{x}^\top \bm{W}^{-1} \bm{y}$ and $s_t^2 = ( \bm{x}^\top \bm{W}^{-1} \bm{x} )^{-1}$ (see \citealp{mandelthesis}, appendix B5), provided that these inverses are available. In the linear regression analogy, $\hat{\phi}_t$ and $s_t^2$ are the posterior mean and variance of $\phi_t$ under a flat improper prior \citep[\S14.7]{gelman_book}. The $\phi_t$-dependent part of Equation \ref{eq:trivariate} can then be simplified into a product of two Gaussian densities, and integrated over $\phi_t$:
\begin{equation}
    \begin{split}
        P(\bm{\phi}_u') &\propto \int_0^\infty \mathrm{ZLTN}( \phi_t | \, \mu_t, \sigma_t^2) \times \mathrm{N}(\phi_t |\, \hat{\phi}_t, s^2_t) \, d\phi_t \\
        &=  \frac{1}{\Psi(\mu_t/\sigma_t)}\int_0^\infty \mathrm{N}(\phi_t | \mu_t, \sigma_t^2) \times \mathrm{N}(\phi_t | \hat{\phi}_t, s^2_t) \, d\phi_t \\
        &= \frac{\mathrm{N}(\mu_t | \, \hat{\phi}_t, \sigma_t^2+s_t^2)}{\Psi(\mu_t/\sigma_t)}\int_0^\infty \mathrm{N}(\phi_t | \, \tilde{\phi}_t, \tilde{\tau}_t^2)  \, d\phi_t \\
        &= \frac{\Psi( \tilde{\phi}_t / \tilde{\tau}_t)}{\Psi(\mu_t/\sigma_t)} \times \mathrm{N}(\mu_t | \, \hat{\phi}_t, \sigma_t^2+s_t^2).
    \end{split}
    \label{eq:phitintegral}
\end{equation}
Here, we have defined for convenience $\tilde{\tau}_t^{2} = [\sigma_t^{-2} + s_t^{-2}]^{-1}$ and $\tilde{\phi}_t = \tilde{\tau}_t^2 (\sigma_t^{-2} \mu_t + s_t^{-2} \hat{\phi}_t)$ when taking the product of the two Gaussian densities. The three steps taken in Equation \ref{eq:phitintegral} correspond closely to those taken in Equations \ref{eq:zltnexpandstep}--\ref{eq:uni_u} when deriving the univariate marginal of $\phi_u^i$. We can reintroduce the terms from Eq.\ \ref{eq:linearmle} that we omitted in Eq.\ \ref{eq:phitintegral} to write the full marginal density of $\bm{\phi}_u'$ as
\begin{equation}
    P(\bm{\phi}_u') = \sqrt{2\pi s_t^2}\frac{\Psi( \tilde{\phi}_t / \tilde{\tau}_t)}{\Psi(\mu_t/\sigma_t)} \times \mathrm{N}(\mu_t | \, \hat{\phi}_t, \sigma_t^2+s_t^2) \times \mathrm{N}(\bm{y} | \, \bm{x} \hat{\phi}_t, \bm{W}),
\end{equation}
where it is important to remember that $\bm{y}$, $\hat{\phi}_t$, and $\tilde{\phi}_t$ are all functions of the two untruncated variables of interest  $\bm{\phi}_u' = (\phi_u^i, \phi_u^j)^\top$.

In the special case of $\bm{\Sigma}_{ut}' = \bm{0}$, the marginalization over $\phi_t$ simplifies considerably:
\begin{equation}
P(\bm{\phi}_u') = N(\bm{\phi}_u' | \bm{\mu}_u', \bm{\Sigma}_{uu}'),
\end{equation}
where $\bm{\Sigma}_{uu}'$ is the $[i,j],[i,j]$ $2\times2$ submatrix of $\bm{\Sigma}_{uu}$.

\bsp	
\label{lastpage}
\end{document}